\documentclass[letterpaper]{article}
\pdfoutput=1
\usepackage{jheppub}
\usepackage{amsmath}
\usepackage{graphicx}
\usepackage[config=altsf]{subfig}
\usepackage{array}
\usepackage{slashed}
\usepackage{afterpage}
\usepackage{appendix}
\usepackage{multirow}
\usepackage{float}
\usepackage{mathtools}
\mathtoolsset{centercolon}

\usepackage{multirow}

\newcommand{\TeV}{{\rm TeV}}
\newcommand{\GeV}{\ {\rm GeV}}

\newcommand{\uv}{{\rm uv}}
\newcommand{\ir}{{\rm ir}}
\newcommand{\KK}{{\rm KK}}
\newcommand{\ew}{{\rm ew}}
\newcommand{\QCD}{{\rm QCD}}

\newcommand{\stil}{S}

\title{Avoided Deconfinement in Randall-Sundrum Models}
\author{Prateek Agrawal}
\author{and Michael Nee}

\affiliation{Rudolf Peierls Centre for Theoretical Physics, 
University of Oxford, Parks Road, Oxford OX1 3PU, United Kingdom}
\emailAdd{prateek.agrawal@physics.ox.ac.uk}
\emailAdd{michael.nee@physics.ox.ac.uk}

\abstract{ We study first order phase transitions in Randall-Sundrum
  models in the early universe dual to confinement in large-$N$ gauge
  theories.  The transition rate to the confined phase is suppressed
  by a factor $\exp(-N^2)$, and may not complete for $N \gg 1$,
  instead leading to an eternally inflating phase. To avoid this fate,
  the resulting constraint on $N$ makes the RS effective field theory
  only marginally under control. We present a mechanism where the IR
  brane remains stabilized at very high temperature, so that the
  theory stays in the confined phase at all times after inflation and
  reheating. We call this mechanism avoided deconfinement. The
  mechanism involves adding new scalar fields on the IR brane which
  provide a stablilizing contribution to the radion
  potential at finite temperature, in a spirit
  similar to Weinberg's symmetry non-restoration mechanism. Avoided
  deconfinement allows for a viable cosmology for theories with
  parametrically large $N$. Early universe cosmological phenomena such
  as WIMP freeze-out, axion abundance, baryogenesis, phase
  transitions, and gravitational wave signatures are qualitatively
  modified.
}
  \preprint{\today}

\begin{document}

\maketitle

\section{Introduction}

Large-$N$ gauge theories are important both theoretically as well as
phenomenologically. The {large-$N$} limit makes many problems in
strongly-coupled gauge theories tractable, providing an expansion
parameter for non-Abelian theories~\cite{Hooft1974a}. It remarkably
also plays a central role in gauge-gravity
duality~\cite{Maldacena1999, Witten1998k,Gubser1998} where a large-$N$
gauge theory is dual to a gravitational theory in higher dimensions.
A major phenomenological application of the duality appears in the
context of the Randall-Sundrum (RS) model~\cite{Randall1999}. 
The RS model provides an elegant solution to the hierarchy
problem, with an exponential separation of the Planck and weak scales
that is generated by the warping of the extra dimension. The ratio of
the two scales is set by the size of the extra dimension, which is
fixed by introducing a stabilisation mechanism,
generating a potential for the radion~\cite{Goldberger1999}. This
mechanism is dual to adding an almost-marginal operator in the gauge theory
which explicitly breaks the scale invariance of the theory. The small
anomalous dimension for this operator, equivalent to a small bulk mass
for the stabilising field~\cite{Witten1998k}, generates an
exponentially small scale in the IR in a manner analogous to
dimensional transmutation in QCD.

The RS model also provides an effective description for a number of
string constructions of realistic vacua with reduced
supersymmetry~\cite{Luty:2000ec, Verlinde2000, Chan2000, Brummer2006, Randall2019}.
In the large-$N$ limit, these explicit constructions are described by 
an effective quantum field theory where gravity is weakly coupled.
A prominent example is the KKLT
scenario~\cite{Kachru:2003aw}, which is partly based on the
Klebanov-Witten~\cite{Klebanov:1998hh} and
Klebanov-Strassler~\cite{Klebanov2000, Klebanov:2000hb} constructions.
For this construction the stability of the de Sitter vacuum is often
justified in the probe approximation which is valid for parametrically
large $N$~\cite{Bena:2018fqc,Kachru:2019dvo}.
Therefore, large-$N$ gauge theories with gravitational duals are
expected to be an important part of a UV complete description of our
universe.

Theories which are described by the RS model in an appropriate limit
suffer from a severe cosmological constraint. At high temperature the
gauge theory is in the deconfined phase, and 
the confined phase becomes thermodynamically preferred
below a critical temperature.
However, the confinement phase transition is first order and
exponentially suppressed with a rate proportional to $\exp(-N^2/
\lambda)$, where $\lambda$ denotes a possible weak coupling.
The gravitational description of the deconfined
phase is the
AdS-Schwarzschild (AdS-S) solution with a UV brane.  The confinement
transition corresponds to the appearance of the IR brane from behind
the AdS-S horizon~\cite{Hawking1983,Witten1998,Creminelli2002}.
The confinement
scale in the gauge theory is dual to the vacuum expectation value of
the radion field which sets the size of the extra dimension in the RS
model.

For large-$N$, the suppressed phase transition is much slower than the
expansion rate of the universe, leading to eternal inflation.
Requiring the phase transition to complete leads to a robust upper bound
on $N$~\cite{Kaplan2006}:
\begin{align}
  N
  \lesssim
  \sqrt{4\lambda \log \frac{M_{\rm pl}}{\Lambda_c}}
  \sim
  12 \sqrt{\lambda}
   \, .
  \label{eq:Nbound}
\end{align}
where $\Lambda_c\simeq1\ \TeV$ is the confinement scale for the gauge
theory. This
bound follows just from dimensional analysis and is independent of the
details of the RS model.  The
ratio $N/4\pi$ sets the hierarchy between the curvature scale $k$ and the
5d reduced Planck scale $M_5$, $N/4\pi \sim (M_5/k)^{3/2}$~\cite{Gubser2001}.  A small value of
this ratio, as
implied by the bound~\eqref{eq:Nbound}, then means that corrections
due to Planck scale physics become important, making the EFT control
in the RS model delicate. 
Gravitational loop corrections can be estimated by the following loop
counting parameter,
\begin{align}
  \frac{N_{\rm species} k^3}{16\pi^2 M_5^3} < 1
  \Rightarrow
  N^2 \gtrsim N_{\rm species}
\end{align}
which is in tension with equation~\eqref{eq:Nbound} even with just the SM
degrees of freedom contributing to $N_{\rm species}\sim 100$.

In fact, the bound is much more stringent
within the simplest version of the RS model.
In this setup, the backreaction of the stabilization mechanism and
breaking of scale invariance are
assumed to be small even close to the confinement scale.  The
gauge theory is an approximately conformal field theory (CFT), with 
spontaneously broken conformal invariance in the confined phase. The
approximate conformality suppresses the phase transition further,
making $\lambda\ll 1$,
so that the bound on $N$ in equation~\eqref{eq:Nbound} is impossible
to satisfy.

There is a large body of work devoted to relaxing this more stringent
constraint on the RS model by changing the details of the
stabilisation mechanism~\cite{Randall2006, Nardini2007,
  Konstandin2010, Konstandin2011, Dillon2017, VonHarling2017,
  Bruggisser2018a, Bruggisser2018, Megias2018a, Bunk2018,
Baratella2019, Agashe2019, Fujikura2019, Azatov:2020nbe, Megias2020, Agashe2020} in
such a way that $\lambda\simeq 1$ and the phase transition occurs more
rapidly.  However, the $N^2$
dependence of the tunnelling rate is a generic
feature of the confinement phase transition. While
modifying the stabilisation mechanism can change the numerical value
of the bound on $N$, in all these models the phase transition is
exponentially suppressed at large-$N$ and therefore subject to the bound in
equation~\eqref{eq:Nbound}.
As pointed out in~\cite{Hassanain2007},
for Klebanov-Strassler type constructions, the effective value of $N$
itself varies over the extra-dimensional coordinate -- the relevant
$N$ in this case is the value near the confinement scale.

In this paper we present a simple modification to the RS model where
$N$ can be made parametrically large without running into this
cosmological bound. We construct a scenario where the confinement
scale grows with temperature, and hence the universe can remain in the
confined phase at all times in early cosmology. For this reason, we
call our mechanism avoided deconfinement.  In order to achieve this we
consider the RS I model with the IR brane stabilised by a
Goldberger-Wise (GW) field $\Phi$~\cite{Goldberger1999}. By
introducing new scalars to the IR brane,
we can generate a potential
which stabilises the
IR brane at high temperatures. The mechanism we use to achieve
this is reminiscent of non-restoration of electroweak symmetry at high
temperatures, as considered in refs~\cite{Weinberg1974, Meade2018,
Baldes:2018nel, Glioti:2018roy, Matsedonskyi2020}.  Similar mechanisms
have also been proposed to avoid monopoles~\cite{Langacker1980, 
Salomonson1985, Dvali1995} or domain walls~\cite{Dvali1995a} in
Grand Unified theories, as models of CP violation at high
temperature~\cite{Mohapatra1979, Mohapatra1979a}, and in the $O(N)
\times O(N)$ models of~\cite{Orloff1996, Chai2020}.

\begin{figure}[tp]
\begin{centering}
\begin{tabular}{cc}
  \includegraphics[width=0.45\textwidth]{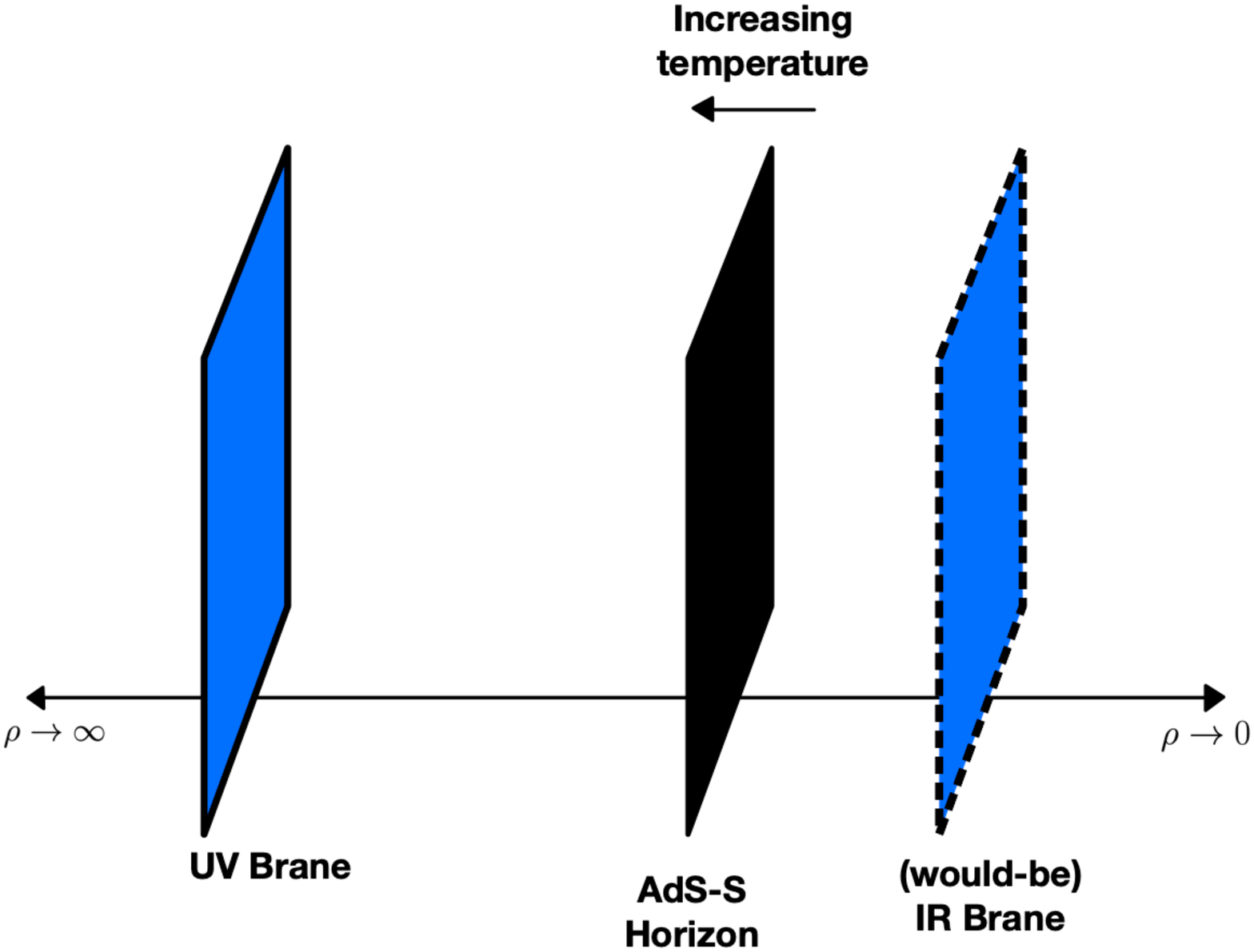}
\hspace{10mm}
 &  \includegraphics[width=0.45 \textwidth]{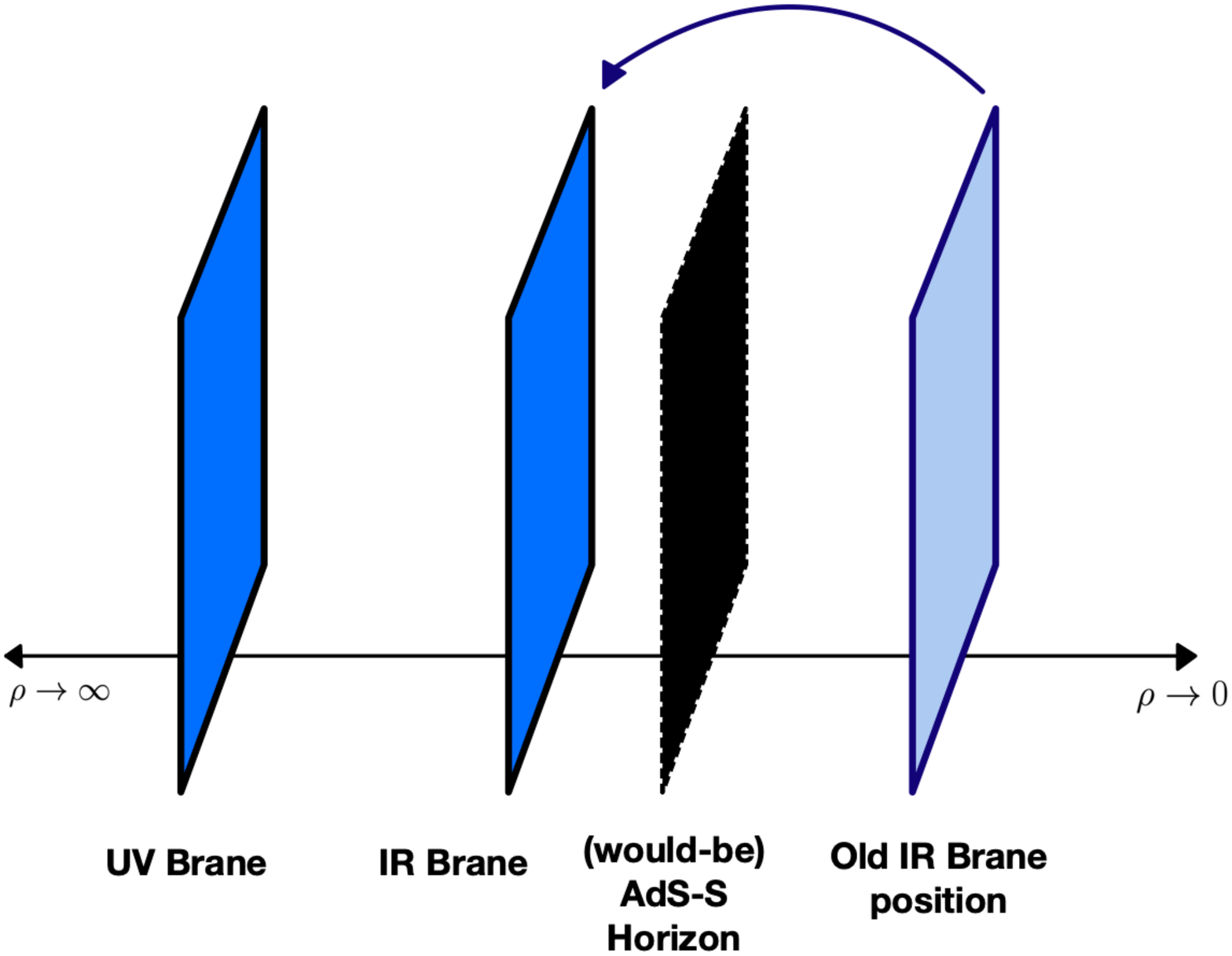}
  \end{tabular}
\end{centering}

\caption{Left hand diagram shows the high temperature behaviour of the
RS model, while the right hand diagram shows the high temperature
behaviour in the AD model. Beyond a certain temperature in the RS
model the IR brane is stabilised behind the location of the horizon in
the AdS-S phase, indicating that the model is unstable against black
hole formation. In the AD model, this instability is lifted by
introducing a temperature dependence to the stabilisation mechanism in
such a way that the IR brane is stabilised outside the would-be
horizon at high temperatures.}
\label{fig:ADdiagram}
\end{figure}

The modification we make to the RS model can lead to dramatic
departures from its standard cosmological history.
Above a critical temperature $T_c$, the confinement
scale varies almost linearly with temperature $T$ leading to
$T$-dependent mass scales on the IR brane,
\begin{align}
  M_{\ir}(T)
  \propto
  \mu(T) = \mu (0) \left(\frac{T}{c T_c}\right)^{1/(1+\epsilon)}
  \label{eq:mirT}
\end{align}
with $\epsilon\ll1, c\sim\mathcal{O}(1)$. 
For a mass scale $M_{\ir} > T_c$, this can imply that
the ratio $T/M_{\ir}(T)$ reaches 1 at very high
temperatures, or potentially not at all (similar to low reheating
temperature models).
Taking the standard model (SM) to be
localised on the IR brane, the $T$-dependence of the electroweak and
QCD scales is as in equation~\eqref{eq:mirT}.
If $v_{\rm ew} > T_c$, the electroweak phase transition occurs
at temperatures far above the TeV scale or is completely avoided.
Furthermore, at
high temperature, fields localised in the UV of the RS model may have
had significant overlap with fields localised towards the IR of the
theory, a feature which may have applications to models of
baryogenesis and dark matter production.

The initial condition for our mechanism to work is that the universe
exits inflation in the RS phase with a stabilized IR brane. A
relatively simple
way to achieve this is to have inflation with Hubble rate below the
confinement scale of the gauge group, or with an additional
stabilization of the IR brane during inflation. 
It will be an interesting future direction to study the interplay of
AD with inflationary models.
After
inflation the universe reheats and the AD mechanism prevents the brane
from falling behind the would-be AdS-S horizon (see
figure~\ref{fig:ADdiagram}).
Note that at high enough
temperatures the AdS-S phase will still be the preferred
thermodynamic phase of the
theory, but in
the avoided deconfinement model the RS phase is classically stable.
The probability of tunnelling from the RS to the AdS-S
phase is exponentially suppressed by
$N^2$ factors, and can be made vanishingly small in the large-$N$
limit.

The rest of this paper is organised as follows. In
section~\ref{sec:Supercooled} we describe the early universe cosmology
and summarise the details of the confinement phase transition in
various generalizations of the RS model that have been considered in
the literature.  We go on to describe the avoided deconfinement (AD)
model in section~\ref{sec:ADmodel} and show how the model leads to a
stabilised IR brane at high temperatures. In
section~\ref{sec:ZeroTPheno} we present the low energy effective
Lagrangian and discuss some of the experimental signatures of the
model. In section~\ref{sec:Cosmology} we then discuss the unique early
universe cosmology of the model and how this relates to other
non-standard cosmological histories in the literature. We also discuss
potential applications of the model to baryogenesis, dark matter
production and potential gravitational wave signatures in this
section, before concluding and summarising in
section~\ref{sec:Discussion}.

\section{The Supercooled Randall-Sundrum Model}

\label{sec:Supercooled}

In this section we review the standard cosmology of the RS type I
model and Goldberger-Wise field, and its dual gauge theory description
via the gauge-gravity duality.
In the standard treatment of gauge-gravity duality at finite
temperature,
the gauge theory partition function is defined
on a manifold $\mathcal{M} = S_1 \times R^3$ with the temporal direction
compactified on a circle of radius $\beta = \pi/T$.
The corresponding gravitational theory is
defined on a 5-dimensional manifold with $\mathcal{M}$ as the boundary.
In computing the gravitational
partition function, all possible geometries $\Sigma$ which satisfy the
boundary condition $\partial \Sigma =  \mathcal{M}$ must be integrated
over~\cite{Witten1998}. The partition function will however be
dominated by classical gravity solutions.
Each
semi-classical gravitational solution $\Sigma_i$ which satisfies
the boundary condition is interpreted as a different phase of the CFT.
At a given temperature, the geometry which minimises the Euclidean
action will give the dominant contribution to the partition function,
and therefore correspond to the preferred phase of the CFT. 

In the RS model, the UV brane cuts off the AdS space, and hence plays
the role of the boundary $\partial \Sigma$. The dual gauge theory is
interpreted as a field theory coupled to 4D gravity, defined
on the manifold $\mathcal{M}$. 
One of the possible classical solutions is,
\begin{align}
  ds^2_{\rm RS}
  &=
  k^2 \rho^2 dt^2
  -\frac{ d\rho^2}{k^2 \rho^2}
  -\rho^2 k^2 dx_i^2,
  \,
  \label{eq:RSmetric}
\end{align}
with the space in the $\rho$ direction cut off at the position of the
IR and UV branes so that $\rho_\ir < \rho< \rho_\uv$.
Here and
throughout this paper, we work in a frame where $\rho_\uv$ is fixed to the
value $\rho_\uv = k^{-1}$, where $k$ is the AdS curvature.
A convenient definition
of the temperature of the 5D theory is the local temperature at the UV
brane. We will simply refer to this temperature as $T$. 
Thermal effects tend to push the IR brane
towards the horizon, rendering the RS solution
unstable~\cite{Creminelli2002} at arbitrarily small temperatures in
the absence of stabilization. 
This instability can be lifted using the GW
mechanism~\cite{Goldberger1999}.

The quasi-conformal theory dual to the RS model is a strongly coupled gauge
theory~\cite{Arkani-Hamed2001, Rattazzi2001} with $\mathcal{O}(N^2)$
degrees of freedom, where $N$ can be determined by matching the
entropy of the black hole with the entropy of the high temperature
phase of the gauge theory~\cite{Gubser2001},
\begin{align}
  \frac{N^2}{16\pi^2} \simeq 12 \left(\frac{M_5}{k}\right)^3.
  \label{eq:Nsquared0}
\end{align}
This relation can be modified by $\mathcal{O}(1)$ factors depending on
strong coupling effects in different gauge theory models.
We see that the large-$N$ aspect of the 4D gauge theory is a
crucial feature of
these models, since it corresponds to the hierarchy between the
curvature scale $k$ and the 5D Planck scale $M_5$
in the 5D gravitational theory. 
The $\rho$ direction can be thought of
as the RG scale of the conformal theory, with small $\rho$
corresponding to the IR of the theory. 
The UV and IR branes of the RS
model correspond to UV and IR cut-offs in the gauge theory.
The cutoff at the IR brane represents a
spontaneous breaking of 
conformality in the IR due to confinement in the gauge theory, while the
UV brane represents explicit breaking by the cutoff at the Planck
scale~\cite{Gubser2001}. The RS model with the IR brane therefore
corresponds to the confined phase of the conformal theory. 
The GW mechanism corresponds
to introducing a
nearly marginal operator to the CFT which explicitly breaks the
conformal symmetry of the theory. The coupling of this operator is
dual to a scalar field in the RS model with a small bulk mass.
Introducing the GW scalar generates an effective potential for the
radion (identified with $ \mu =k^2 \rho_{\ir}$ in co-ordinates where
  the location of the 
UV brane is fixed~\cite{Charmousis2000}), with a minimum at small $\mu$
--  the IR brane will then be stabilised at the minimum of this
potential.

The RS solution with the IR brane becomes classically unstable at high
temperatures. There is another classical solution that contributes to
the finite temperature partition function given by the
AdS-Schwarzschild (AdS-S) geometry, 
\begin{align}
  ds^2_{\rm AdS-S}
  &=
  f(\rho) dt^2
  -\frac{ d\rho^2}{f(\rho)}  -\rho^2 k^2 dx_i^2
  ,\quad
  f(\rho) = k^2\left(\rho^2- \frac{\rho_h^4}{\rho^2}\right)
  \,.
  \label{eq:AdS-Smetric}
\end{align}
The position of the
horizon $\rho_h$ is set by the temperature $\rho_h = \pi T/k^2$. The
solution is cut off at $\rho =\rho_{\uv}$ by the UV brane as before.
The AdS-S solution is dual to
the deconfined phase of the gauge theory, with the Hawking temperature
and entropy of the AdS black hole equal to the corresponding quantities
in the gauge theory. The AdS-S solution is classically stable for any
non-zero temperature, and is the thermodynamically preferred phase of
the theory at high temperatures.

As the universe cools below a critical temperature, the RS phase with
the IR brane becomes preferred and
there is a first order phase transition between the two phases which
proceeds through a tunnelling process connecting the
two solutions. This tunnelling process is strongly suppressed,
however, due to the large change in free-energy in the two phases. The
requirement that this phase transition completes places bounds on $N$
for the model to be cosmologically viable. These bounds typically
require $N\sim \mathcal{O}(1)$, which is in tension with the
assumption of working in the large-$N$ limit.

\subsection{(De)confinement Phase Transition in the
RS Model}
\label{ssec:deconfRS}
We consider the RS model with the GW stabilization mechanism.
The bulk Lagrangian contains gravity and the GW field ($\Phi$),
\begin{align}
  S_{\rm bulk, RS} [G_{AB}, \Phi ] 
  &=
  \int d^4 x \, d \rho \, \sqrt{G} 
  \left[ 
    -2 M_5^3 R+ 24 M_5^3 k^2	
    + \frac12 G^{AB} \partial_A\Phi\partial_B \Phi
    - \frac12 m_\Phi^2  \Phi^2 
  \right]
  \,.
  \label{eq:RSbulk}
\end{align}
We also include brane localized terms,
\begin{align}
  S_{\uv,\rm RS} 
  &= 
  \int d^4 x \sqrt{-g_{\uv}} 
  \left[
    -\lambda_{\uv} (\Phi^2 - v_{\uv}^2)^2 
    -24 M_5^3 k +\delta \Lambda_{\uv}
  \right]
  \label{eq:RSUV}
  \\
  S_{\ir,\rm RS} 
  &=
  \int d^4 x \sqrt{-g_{\rm ir}} 
  \left[
    -\lambda_{\ir} (\Phi^2 - v_{\ir}^2)^2 
    + 24 M_5^3 k + \delta \Lambda_{\ir}
  \right]
  \,.
  \label{eq:RSbraneterms}
\end{align}
where $g_{\uv, \ir}$ are the induced metrics on the UV and IR branes.
In the presence of the GW stabilization mechanism,
only one combination of the brane tension detuning parameters 
$\delta \Lambda_i$ needs to be tuned,
corresponding to the tuning
of the 4D cosmological constant.
Depending on the sign of $m_\Phi^2$, these parameters may be required
to lie in a
certain range for there to be a local minimum for the radion away from
$\mu =0$. 
For simplicity, here we set each of the detuning parameters
 to 0 and assume $m_\Phi^2 > 0$. 
We also assume that the stabilization occurs in the limit of small
backreaction. These assumptions are not crucial and do not affect the
qualitative results. With these assumptions, the metrics
in equations~\eqref{eq:RSmetric} and \eqref{eq:AdS-Smetric} continue
to be approximate classical solutions.

We can obtain the potential for the radion by integrating over the
classical solution for the GW field.
In the limit where the brane localized terms fix
$\Phi(\rho_{\uv,\ir}) = v_{\uv,\ir}$, the 4D effective potential
for the radion, $\mu  \equiv k^2 \rho_{\ir}$ is~\cite{Creminelli2002},
\begin{align}
  V(\mu)
  &=
  \epsilon v_{\uv}^2 k^4
  +\left[
    (4+2\epsilon)\mu^4(v_{\ir} - v_{\uv} (\mu/k)^{\epsilon})^2
    -\epsilon v_{\ir}^2\mu^4
  \right]
  +\mathcal{O}(\mu^8/k^4)
  \, ,
  \label{eq:radion-potential}
\end{align}
with $\epsilon = \sqrt{4 + m_\Phi^2 /k^2} - 2$.
The minimum is obtained for:
\begin{align}
  \mu_{\TeV}
  &=
  f(\epsilon)
  k \left(\frac{v_\ir}{v_\uv} \right)^{1/\epsilon}
  \,,
  \label{eq:RSmumin}		\\
 f(\epsilon) 
 &=
 \left[\frac{4+\epsilon+\sqrt{\epsilon(4+\epsilon)}}
 {4+2\epsilon}\right]^{1/\epsilon}
 \sim \mathcal{O}(1)
 \,.
 \label{eq:functionf}
\end{align}
A relatively modest hierarchy in $v_{\uv,\ir}$ and $\epsilon \sim
1/10$ can generate an exponential hierarchy between $k$ and
$\mu_{\TeV}$.
At energies $\lesssim \mu_\TeV$ the effective theory is a 4D theory
with a tower of Kaluza-Klein (KK) states with masses $\sim
\mu_{\TeV}$.
In the 4D theory, dimensionful parameters involving fields localized on
the IR brane, such as the Higgs mass parameter, scale with
$\mu_{\TeV}$, thus explaining the electroweak hierarchy elegantly.

At low-temperatures,
$T < \mu_\TeV$, both classical solutions -- the stabilized RS solution
with UV and IR branes, and the AdS-S solution with a UV brane and a
black hole horizon -- are (meta)stable.  However, at high-temperatures
$T \gg \mu_{\rm TeV}$, the minimum of the radion potential is behind
the AdS-S horizon
$\mu_\TeV < \rho_h k^2$, indicating
that the AdS-S solution is the only classical solution. 
During the
early universe the universe is in the AdS-S phase; to get to the RS phase
the universe needs to undergo a first order phase transition.

The tunnelling rate per unit volume for the phase transition is,
\begin{align}
  \Gamma &\simeq  R_c^{-4} \exp(-S_b)
\end{align}
where $S_b$ is the bounce action for the tunnelling
transition, and $R_c$ is the radius of the critical bubble~\cite{Coleman1977, Linde1983}. 
The field configuration for the transition from the AdS-S phase to the
RS phase involves moving the black hole horizon to the far IR, $\rho_h
\to 0$, and then nucleating the IR brane at $\rho = 0$ and bringing it
to larger values of $\rho$. 
Therefore this field configuration probes the geometry in the region
where the local temperature is super-Planckian and stringy corrections
would be relevant. However, in the case where the transition
temperature is low, and there is an approximate conformal symmetry in
the IR, the dominant contribution to the bounce is
dictated by the radion dynamics and can be estimated while ignoring
the gravitational contribution to the bounce~\cite{Agashe2020}. 
Even so, since the field configuration probes the
geometry in the far IR, the bounce action for this
configuration can depend sensitively on the
details of the GW stabilisation, and other physics in the IR. 
We summarize the results for the bounce action that have been
considered in various limits in the literature next.

\subsection{Bounce action from the radion}

In a large class of models, the phase transition is captured by the
dynamics of the radion~\cite{Creminelli2002,Agashe2019, 
Agashe2020}.
The general radion effective field theory can be
understood in terms of the dual 4D theory. The 4D theory is a
near-conformal field theory coupled to gravity. The gravitational
sector breaks the conformal
symmetry explicitly, but below the gravitational cutoff an approximate
conformal symmetry survives.
For a stabilised RS geometry with
a light radion, the 4D effective theory below the KK scale $\mu_\TeV$
is well-described by an effective theory of a spontaneously broken
(approximate) conformal symmetry, with the light radion/dilaton as the
pseudo-Nambu-Goldstone boson.

In this section we work in this 4d picture to study a few such
generalisations that have been studied in the literature. As we will
see, in each case the first order phase transition is highly
suppressed. The effective Lagrangian for
the dilaton~\cite{Creminelli2002,Coradeschi2013} can be written as,
\begin{align}
  \mathcal{L}_{\rm eff}
  &=
  \frac{N^2}{16\pi^2} 
  \left [(\partial \mu)^2 - \lambda(g(\mu)) \mu^4\right]
  \,,
  \label{eq:radion-effL}
\end{align}
where the $\mu$-dependence in $g(\mu)$ denotes the explicit breaking
of conformal symmetry due
to the GW deformation. We expect the dilaton to be 
the lightest bound state of the gauge theory~\cite{Goldberger2000,
Pomarol2019} as it is 
the pNGB of the broken dilation symmetry, so is the only relevant
degree of freedom in the IR of the theory. The $N^2$ factor makes
explicit the fact that the dilaton is interpreted as a glueball state
in a 4D large-N gauge theory.

The free energy in the (de)confined phase is well approximated by,
\begin{align}
  F_{\rm confined}
  &= V(\mu_{\TeV}) 
  = -\frac{N^2}{16\pi^2} \lambda_\TeV\mu_\TeV^4 +V_0
  \nonumber\\
  F_{\rm deconfined}
  &= C - 2\pi^4 (M_5/k)^3 T^4
  = C - \frac{\pi^2}{96}N^2 T^4
\end{align}
where we have defined $\lambda_\TeV \equiv |\lambda(g(\mu_\TeV))|$, and
added a constant $V_0$ to ensure that the vacuum energy at the
minimum is zero. Notice that $\lambda(g(\mu_\TeV)) < 0$ for $\mu_\TeV$
to be the minimum of the potential. $C$ can be calculated by matching
the free energy at
$\mu = T = 0$. The critical temperature can be calculated by
equating the free energy
in the two phases at the transition,
\begin{align}
  C-\frac{\pi^2}{96} N^2 T_c^4
  &\simeq
  -\frac{N^2}{16\pi^2} 
  \lambda_\TeV 
  \mu_\TeV^4 +V_0	\\
  \Rightarrow 
  T_c &\simeq \left[\frac{6\lambda_\TeV}{\pi^4}\right]^{1/4} \mu_{\rm TeV}
  \,.
  \label{eq:Tc}
\end{align}
When $\lambda_\TeV \ll 1$, the transition
temperature $T_c \ll \mu_\TeV$, and the approximation of radion
domination is well justified.

If the phase transition is prompt, it completes for $T\sim T_c$. In
this case the bubble has O(3) symmetry, and the action can be estimated in the
thin-wall regime (see e.g.~\cite{Kaplan2006}),
\begin{align}
  \frac{S_3}{T}
  &\sim
  \frac{N^2}{8}
  \left[\frac{1}{\lambda_\TeV}\right]^{3/4}
  \frac{T_c}{T}
  \left( 1-\left( \frac{T}{T_c} \right)^4 \right)^{-2}
  \label{eq:kaplans3}
  \,.
\end{align}
This explicitly shows the general enhancement of the bounce action by
$N^2$, and often also by the weak coupling $\lambda_\TeV$.

We can evaluate the bounce action for the GW model considered in
section~\ref{ssec:deconfRS} above.
The quartic $\lambda_\TeV$ in this case is,
\begin{align}
  \lambda_\TeV
  &=
  \frac{16\pi^2}{N^2 }\epsilon^{3/2} v_\ir^2
  \label{eq:muforGW}
\end{align}
which leads to the following parametric form for the bounce
action~\cite{Creminelli2002},
\begin{align}
  \frac{S_3}{T}
  &\sim
  \frac{N^{7/2}}{\epsilon^{9/8} v_\ir^{3/2}}
  \frac{T_c}{T}
  \left( 1-\left( \frac{T}{T_c} \right)^4 \right)^{-2}
  \,.
  \label{eq:creminells3}
\end{align}
The action is not only enhanced by the factor of $N^2$, but also by
the small quartic coupling of the radion, which increases the
dependence on $N$ to $N^{7/2}$. There is an additional enhancement by
$1/\epsilon$, related to the fact that the scale symmetry violation at
$\mu_\TeV$ is parametrised by $\epsilon$. The exact power of
$\epsilon$ that appears can depend on the implementation
of the GW mechanism~\cite{Creminelli2002,Nardini2007,Agashe2019, Agashe2020}.

More generally, we can see that the action is enhanced for small scale
symmetry violation encoded in $\beta_\lambda \ll 1$. 
The zero temperature minimum is determined by the running quartic,
$\lambda(g(\mu))$, 
\begin{align}
  \partial V(\mu)/\partial \mu = [4\lambda(g(\mu_\TeV))  +
  \beta_\lambda(g(\mu_\TeV))] \mu^3 = 0
  \,.
\end{align}
Thus for a nearly scale invariant theory at $\mu_\TeV$, $\lambda_\TeV$
will be generically small.

If the transition is not prompt, then it will then take place in the
supercooled regime, $T\ll T_c$.
In the case where
there is a barrier in the radion potential between $\mu_{\rm TeV}$ and
$\mu \sim 0$, the bounce configuration is essentially the same as the
zero temperature tunnelling, and has an O(4) symmetric bounce
action~\cite{Creminelli2002, Nardini2007}
\begin{align}
  S_4
  &\sim
  \frac{N^2}{16\pi^2 \lambda_\TeV} .
  \label{eq:S4}
\end{align}
As before, the explicit factor $N^2$ appears. Again, we see that in
the case of the simplest RS+GW model above, the parametric dependences are even
stronger,
\begin{align}
  S_4
  &\sim
  \frac{N^4}{(4\pi)^4 \epsilon^{3/2} v_\ir^2} .
  \label{eq:S4RSGW}
\end{align}

If there is no barrier between $\mu\sim 0$ and $\mu = \mu_\TeV$, then
the ``release point'' for the radion field can be very small $\mu \sim
T$ even for supercooled transition, so the smallest bounce action is
still obtained by an O(3) symmetric bounce. For example, in the case
where conformal symmetry is restored in the IR ($\epsilon > 0$ for the
GW field)~\cite{Agashe2019}, the radion potential near the origin is
$V \sim \lambda(0) \mu^4$. The bounce action is then the $T\ll T_c$
limit of equation~\eqref{eq:kaplans3},
\begin{align}
  \frac{S_3}{T}
  &\sim
\frac{N^2}{8[\lambda(0)]^{3/4}}
  \label{eq:S3ramansupercool}
\end{align}
which is no longer suppressed by the small parameter $\epsilon$. In
the case where $\epsilon \lambda_\TeV$ is not parametrically small,
radion dynamics no longer suffice to estimate the bounce. However, it
may still be possible to estimate the bounce in the 5D effective
theory~\cite{Agashe2020} and is found to be $\mathcal{O}(N^2)$.

When $\epsilon < 0$, the GW field profile grows towards the IR.
Consequently, the higher order terms in the GW potential might become
important and the approximate conformality $\partial_{\log \mu} g(\mu)
\sim \epsilon$ might be broken as we approach $\mu \lesssim \mu_{\rm
TeV}$. In such cases the enhancement of the bounce action by
$1/\epsilon$ will be absent, even though the EW/Planck hierarchy is
set by small epsilon. This can be explicitly seen in explicit
holographic constructions~\cite{Bigazzi2020, Hassanain2007}, or in RS models with more
general stabilisation mechanisms~\cite{Konstandin2010, Nardini2007, Konstandin2011, Dillon2017,VonHarling2017, Bruggisser2018, Bruggisser2018a, Megias2018a, Baratella2019, Agashe2019, Agashe2020, Fujikura2019, Megias2020, Bunk2018}.

We see from the examples above that while the details of the bounce
action depend on the actual theory, it takes the form $S_b \simeq
N^2/\lambda$ in each case, with $\lambda \lesssim 1$. 
This
has far-reaching consequences for early universe cosmology.
Either the universe is required to be reheated to temperatures lower
than the confinement scale, or there is a
strong constraint on the maximal $N$ allowed. 

If the rate of tunnelling is smaller than Hubble,
the universe will get stuck in the false vacuum~\cite{Guth1983}. Since
the true vacuum
at zero temperature is assumed to have a (nearly) vanishing cosmological
constant, the deconfined vacuum  has a large positive
cosmological constant $C \sim N^2 T_c^4$ at low temperatures and
starts to inflate with $H\simeq N T_c^2 / M_{\rm pl}$.
In a Hubble volume, the
probability of completing the phase transition within a Hubble time
is,
\begin{align}
  P
  &= \Gamma/H^4
  \label{eq:tunnel}
\end{align}
If $P \ll 1$, then the universe eternally inflates. This gives us a
bound on $N$,
\begin{align}
  N
  &\lesssim 
  2
\sqrt{
  \lambda \log \frac{M_{\rm pl}}{T_c}}
\end{align}
We have replaced the inverse critical radius by $T_c$; unless the
bubble size is exponentially smaller, this is a reasonable
approximation.
In many models considered above, $\lambda$ is parametrically
small. For the RS+GW model above, $S_b$ is enhanced both
by $N$ as well as $1/\epsilon$, making it impossible to satisfy the
constraint above. 
Even without these enhancements, the calculations above
assume dilaton dominance, which requires $\lambda_\TeV \ll
1$~\cite{Agashe2019}.
Therefore, it is hard to get $S_b \lesssim N^2$ in a controlled
approximation. This translates into a bound $N\lesssim 12$ for $T_c
\sim 1\,\TeV$. 
From equation~\ref{eq:Nsquared0}, we see that the 
hierarchy between the 5D Planck scale and the AdS curvature is 
$(M_5/k)\lesssim 1$.  This lack of hierarchy
makes the 5D effective gravitational theory very delicate. 

One avenue to evade this cosmological bound is to avoid reheating the
universe above the TeV scale. This may require a more intricate
inflationary mechanism, as well as solutions to baryogenesis at the
electroweak scale or below~\cite{Baldes:2018nel, Bruggisser2018, Bruggisser2018a}.
In the next section we outline the avoided deconfinement mechanism,
where the GW stabilisation of the radion is temperature dependent and
the IR brane is stabilised at arbitrarily high temperatures. This
allows for parametrically large hierarchies between $M_5$ and $k$, and
an
early cosmology without a stringent restriction on the reheat
temperature.

\section{5D Model for Avoided Deconfinement}

\label{sec:ADmodel}

In this section we modify the RS model with GW field ($\Phi$) by
including  extra scalars
localised to the IR brane\footnote{Localised fields on the IR brane may be required to arise from corresponding bulk modes with masses below the 5D cutoff~\cite{Fichet:2019owx}. These bulk modes will then have an associated tower of KK states, but this detail will not affect our discussion.}. Given a suitable set of parameters, the
effect of this will be to realise a model where the new scalars
provide a metastable minimum for the radion at high temperature, avoiding
the formation of the AdS-S black hole.

We make a simple modification to the RS model described in
equation~\eqref{eq:RSbraneterms} by
adding scalar field(s) $\stil$ to the IR brane. The action is:
\begin{align}
  S
  &=
  S_{\rm bulk, RS}
  + S_{\uv,\rm RS}
  + S_{\ir,\rm RS}
  + S_{\ir,\rm AD}
  \,.
  \label{eq:ADactiondeff}
\end{align}
where $S_{\rm bulk, RS}$ and $S_{\uv/\ir,\rm RS}$ are the RS model bulk and
brane actions which are unchanged from their definition in
equations~\eqref{eq:RSbulk}, \eqref{eq:RSUV},and~\eqref{eq:RSbraneterms}.
We
continue to choose the detuning parameter $\delta \Lambda_\ir = 0$.
As usual, this
simplifying assumption can be relaxed.
The modified IR brane action, $S_{\ir,\rm AD}$ includes
$N_s$ real scalars $\stil$ localised to the brane.
The additional terms in the IR brane action is
\begin{equation}
\begin{aligned}
  S_{\ir,\rm AD} 
  &= 
  k^4
  \int_{\rho = \rho_\ir} d^4 x \sqrt{-g_\ir}
  \sum_{i=1}^{N_s}
  \left[
    \frac{1}{2 k^2} g_{\ir}^{\mu \nu}
    \partial_\mu \stil_i \, \partial_\nu \stil_i
    - \frac{\lambda_s}{4}( \stil_i^2 - v_s^2)^2
    - \frac{\gamma}{6} \stil_i^3 
  \right]
  ,
  \label{eq:IRaction}
\end{aligned}
\end{equation}
where we have explicitly included factors of $k$ so that the
parameters $\lambda_{s}, \gamma, v_s$ as well as the field $\stil$
are dimensionless. We will suppress the index $i$ on $S$ for
notational simplicity.  In order for the potential to remain bounded from
below, the coupling $\lambda_{s}$ must be positive.  The value of the
masses and quartic couplings of each field $\stil$ do not have to be
equal, but for simplicity we will take them to be the same. For
unequal couplings our results below can be reinterpreted using
statistical averages over the $\stil$ ensemble. Each $\stil$ has an
approximate $Z_2$ symmetry that is spontaneously broken at zero
temperature. The coupling $\gamma$ that weakly breaks the $Z_2$
symmetry for each $\stil$ is introduced to avoid domain wall problems,
and can be very small in a technically natural way.

Before presenting the consequences of adding these extra scalars, we summarize the
choice of parameters for which our approximations are under theoretical control. 
The primary goal of the AD setup is to generate a classically stable minimum for the radion at high temperatures for arbitrary~$N$, therefore avoiding a confinement phase transition entirely and putting the large-$N$ approximation on a firmer footing. It is then worth 
highlighting the validity of the large-$N$ expansion, especially in light of adding extra matter on the IR brane. Requiring the gravitational loop counting parameter to be small,
$N_{\rm species}/N^2 \lesssim 1$, restricts the number of scalars we can add. As we show below, the AD mechanism does require $N_s > 1$ to operate. However, in order to obtain a classically stabilized radion at high temperatures, $N_s$ can be parametrically smaller than $N^2$. In this case, the black hole phase does have a lower free energy, but the tunneling rate from the AD phase to the black hole phase is exponentially suppressed by a tunnelling exponent of order $\sim N^3$. We present an estimate of the tunneling rate in appendix~\ref{app:tunneling}. Thus, the parameter $N$ can be taken arbitrarily large while keeping other parameters in our model fixed, ensuring that the $1/N$ expansion is well under control.

In order to understand the other parametric scalings in the Lagrangian, it is useful to characterize the cosmological history of the AD model by the following three scales:
\begin{enumerate}
	\item the temperature $T_s$ at which the scalars $S$ undergo a (crossover) phase transition;
	\item the temperature $T_c$ at which the AD construction begins to take effect and the position of the IR brane begins to vary with temperature; and
	\item the zero-temperature radion vev, $\mu_\TeV$.
\end{enumerate}
The AD model requires the hierarchy $T_s < T_c$, due to the fact that the addition of the new scalars $\stil$ only generates the desired finite-temperature effects in the symmetric phase.
At temperatures $T>T_c$ the position of the IR brane is moved from the GW minimum due to thermal effects. As mentioned above, in this temperature range the confined phase is metastable, in contrast to the usual RS model where the confined phase becomes classically unstable at high temperature. Another condition on the parameter space comes from the requirement that the IR brane is stabilized at a radius where the local temperature is small enough such that the backreaction on the bulk geometry is small. At the confinement scale, this condition implies
\begin{align}
 T_c< \frac{\mu_\TeV}{\pi},
\end{align}
This condition becomes stronger logarithmically in temperature, and can fail at very high temperatures, which sets a maximum temperature $T_{\rm max}$ for the AD mechanism to operate.
 As we explain in further detail in section~\ref{sec:HighTstabilization}, this leads to the following condition on the parameters of the model:
\begin{align}
  1 > \frac{\pi T_c}{\mu_\TeV} > \sqrt{\frac{6}{N_s}} \frac{m_\varphi}{m_s} > v_s
  \label{eq:param-inequalities}
\end{align}
where $m_\varphi, m_s$ are the masses of the radion and scalar ($s$) fluctuations at zero temperature. Since the mass of the radion $m_\varphi$ cannot be too small phenomenologically, this leads us to require a moderately large number of scalar fields $N_s$ for the AD model to work.

\subsection{Finite temperature effective potential for the radion}
\label{sec:IRPotential}

We work in a regime where the local 5D temperature remains below
$(M_5^3 k^2)^{1/5}$ everywhere in the 5th dimension, so that the
finite temperature effects have a
negligible backreaction effect on the bulk geometry.
At any temperature $T$ and the position of the IR brane $\rho_\ir(T)$,
we can solve the equations of motion for the GW field on the
background RS metric with the same
boundary conditions, $ \Phi(\rho_{\ir / \uv}(T))=k^{3/2} v_{\ir / \uv}$.
The bulk solution is,
\begin{align}
  \Phi(\rho) = A \rho^{-4-\epsilon} + B\rho^{\epsilon}
\end{align}
where $A,B$ are fixed by the boundary conditions.
As above, $\epsilon \approx m_\Phi^2/(4k^2)$, which we take to be positive.
We expand in fluctuations around this classical solution
with the size of the extra dimension equal to $\rho_\ir(T)$.
We decompose the bulk fluctuations into Kaluza-Klein modes and
integrate over the 4D modes to derive the finite temperature
effective potential.

The temperature-dependent effective potential can be broken up into
the tree-level potential and the one-loop potential~\cite{Delaunay2008, Curtin2018}:
\begin{align}
  V_{\rm eff}(T)
  &=
  V_{\rm tree}
  +
  \Delta V^{\rm CW}_{1}
  +
  \Delta V^T_1(T)
  \label{eq:potential-split}
\end{align}
where we have separated the 1-loop contribution into a piece that goes to
0 at zero temperature $\Delta V^T_1(0) = 0$. The zero temperature
Coleman-Weinberg potential, $\Delta V^{\rm CW}_1$ includes the
usual UV-divergences one would encounter in these calculations. 
The tree-level 4D action is obtained as above by integrating the classical
solution over the extra dimension. The potential is given by,
\begin{align}
  V_{\rm tree}(\mu , \stil_i)
  &=
  \mu^4
  \left[
    (4+2\epsilon) (v_\ir-v_\uv (\mu/k)^\epsilon)^2
    -\epsilon v_\ir^2 
    +\sum_{i=1}^{N_s}
    \left(
    \frac{\lambda_s}{4}(\stil^2_i-v_s^2)^2 
    +\frac{\gamma}{6} \stil_i^3 
  \right)
  \right]
  \label{eq:VtreeT}
\end{align}
where we have used $\mu = k^2 \rho_\ir(T)$. We have also suppressed the
$T$-dependence in the notation.

The one-loop contribution is obtained by integrating over the
fluctuations.
The finite temperature contribution to the potential depends on the
effective mass of the fluctuations around the classical solution. The
relevant particles in our case are the radion, the new scalars and
the SM fields. 
Including the kinetic term for the radion~\cite{Coradeschi2013,
Chacko:2014pqa} and the scalars, 
we find the following action for the fluctuations:
\begin{align}
  S
  &= 
  \int d^4x 
  \left[ 
    \mathcal{L}_{\rm SM}
    +\frac12(\partial \varphi)^2 
    +\frac12( \partial s_i)^2
    - V_{\rm tree} 
    \left(\mu \left(1 + \frac{\varphi}{F_{\varphi}}\right),
      \stil_i + \frac{s_i}{\mu}\right) \right]
\label{eq:radionaction} 
\end{align}
We have introduced the canonically normalized fluctuations for the
radion $\varphi$ and the
scalars $s$. The radion decay constant 
\begin{align}
  F_\varphi = \frac{N}{2\sqrt{2}
  \pi} \mu.
  \label{eq:Fphi}
\end{align}
The indices in the kinetic term are now contracted with the 4d Minkowski metric.

The field-dependent masses of these particles are defined as the
second derivative of the potential w.r.t.~the corresponding
field. The SM particle masses and the radion and $s$ masses all scale
with  $\mu$. For a large number of scalars $N_s \gg 1$, the
thermal potential is dominated by loops of $s_i$\footnote{Note that
  since the SM contribution to the $\mu$ potential is proportional the
mass of the SM field, only the contributions from $t,W,Z,h$ are
sizeable.}. The field-dependent masses of $s_i$ are,
\begin{align}
  m_{s,i}^2
  &=
\mu^2
\left[-\lambda_s v_s^2 + 3 \lambda_s \stil_i^2 + \gamma \stil_i\right]
  \label{eq:fluctuation-masses}
\end{align}
The other modes in the
spectrum are the KK modes of the graviton, the GW field and other
fields in the bulk.
The mass of the $n$th 
KK mode is approximated by~\cite{Gherghetta2000}:
\begin{align}
  m_n 
  &\simeq 
  \left( n + \frac{2+\epsilon}{2} -\frac{3}{4} \right) \pi \mu .
\end{align}
The $\rho$-coordinate of the would-be AdS-S horizon is $\rho_h = \pi
T/k^2$. Therefore, the condition that the IR brane is stabilised
outside the AdS-S horizon, $\rho_\ir(T) > \rho_h$ implies that the
higher
KK modes are not excited at any $T$, and can be safely
neglected in the thermal potential.

The Coleman-Weinberg potential is given by,
\begin{align}
  V_1^{\rm CW} (\mu, S_i)
  &=
  \sum_{i=1}^{N_s}
  \frac{1}{64\pi^2}
  m_{s,i}^4 \left( \log\left[\frac{m_{s,i}^2}{\mu_R^2}\right] -\frac32\right)
\end{align}
where $\mu_R$ is a renormalisation scale. 
A convenient choice of the
renormalisation scale for the dynamics on the IR brane is $\mu$
itself~\cite{Sundrum:2003yt}. With this choice, we do
not generate large hierarchies of scale
on the IR brane and the one-loop corrections at zero-temperature stay small.
We have included all terms allowed by the scale symmetry of the radion
and the $Z_2$ symmetries in the $s$-sector, as well as leading terms
violating these symmetries parameterized by the small parameters
$\epsilon,\gamma$. Thus, the higher order terms can be safely
neglected and we will simply absorb the UV divergent pieces into a
redefinition of couplings and masses as renormalized quantities.

The finite-temperature one-loop contributions from $s_i$ are
\begin{align}
  \Delta V_1^T(\mu(T),\stil_i(T),T)
  &=
  \sum_{i =1}^{N_s}
  \frac{T^4}{2\pi^2}
  \int dk\, k^2 
    \log\left[
      1 - \exp\left(-
        \sqrt{
    k^2 + \frac{m_{s,i}^2(\mu,\stil_i)}{T^2}
  }
\right)
  \right]
  \\&
  \equiv
  \sum_{i =1}^{N_s}
  \frac{T^4}{2\pi^2}
  J_b\left(\frac{m_{s,i}^2(\mu, \stil)}{T^2}\right)
  \label{eq:V1T}
\end{align}
We approximate the thermal function $J_b$ by assuming that
$m_s(T) \ll T$.
At high temperature the thermal function can be approximated as,
\begin{align}
  J_b(y^2)
  &\approx
  -\frac{\pi^4}{45}
  +\frac{\pi^2}{12} y^2
  -\frac{\pi}{6} y^3
  -\frac{1}{32} y^4 \left(\log \frac{y^2}{\pi^2} +2 \gamma_E -
  \frac32\right)
  \,.
  \label{eq:thermal-function}
\end{align}
where $\gamma_E$ is the Euler-Mascheroni constant.
The field-dependent $\log$ pieces cancel between the Coleman-Weinberg
terms and the thermal corrections. The renormalisation scale $\mu_R
\simeq \mu$ scales with the temperature (as we show below), and hence
 we do not
get any enhanced large-log pieces, and we can safely ignore the terms
of $\mathcal{O}(y^4)$. 
Then,
\begin{align}
  \Delta V_1^T(\mu,S,T)
  &\simeq
  \frac{N_s}{24}
  T^2 \mu^2
  \left(
    -\lambda_s v_s^2 
    + 3 \lambda_s {\stil^2}
    + \gamma {\stil}
  \right)
  \nonumber\\
  &\qquad
  -\frac{N_s}{12\pi}
  T \mu^3
  \left(
    -\lambda_s v_s^2 
    + 3 \lambda_s {\stil^2}
    + \gamma {\stil}
    +\lambda_s \frac{T^2}{4\mu^2}
  \right)^{3/2}
  \label{eq:V1T2}
\end{align}
where we have used the fact that each of the scalar vevs $S_i = S$
when $\{\lambda_s, v_s, \gamma\}$ are taken to be the same for each $s_i$. 
The extra term involving $T^2$ in the second term above is a result of
performing the leading daisy resummation, where we replace the
field-dependent mass $m_{s,i}^2$ by $m_{s,i}^2 + \Pi_i$ in
equation~\eqref{eq:V1T}, with $\Pi_i$
the leading temperature contribution to the one-loop thermal mass.

Thus, we see that at high temperatures $T > T_s \simeq \frac12v_s \mu_\TeV$, thermal effects drive to
restore the (approximate) $Z_2$ symmetry in $\stil$, so that $\langle
S \rangle \ll 1$. This generates a
tachyonic direction for $\mu$, providing a finite temperature
stabilization. As the universe cools, the $S$ symmetry gets broken, and
the radion settles down close to its zero temperature minimum dictated
by the GW part of the potential.
The thermal potential can be minimized numerically,
and for a range of parameters the radion remains stabilized outside
the would-be AdS-S horizon at high
temperatures.

\subsection{High temperature radion stabilization}
\label{sec:HighTstabilization}

\begin{figure}[tp]
\begin{centering}
  \begin{tabular}{ll}
      \hspace{-10mm}
	\includegraphics[width=0.5\textwidth]{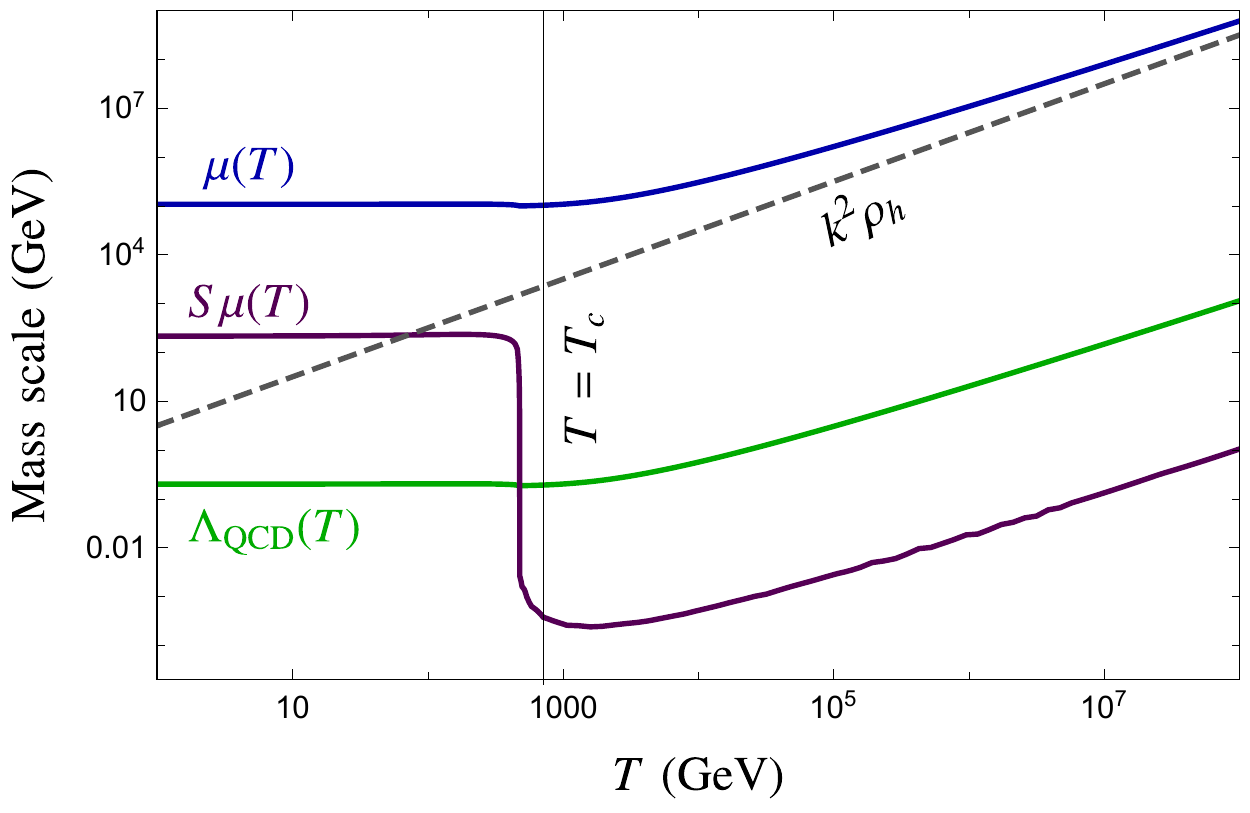}
	&
	\includegraphics[width=0.5\textwidth]{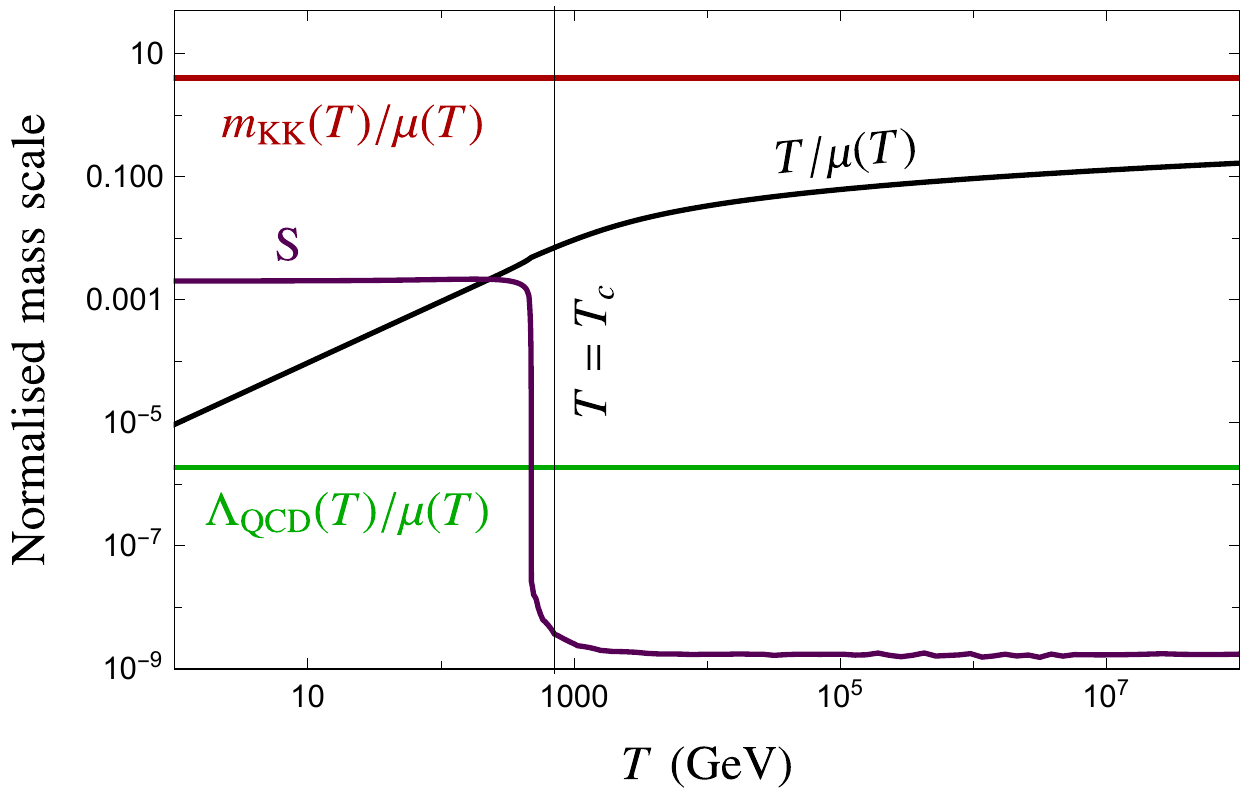}

	  \end{tabular}
        \caption{ Plots showing the dependence of mass scales in the
        theory with temperature, for parameters in equation
      \eqref{eq:num-params}. The left plot shows the radion
    expectation value ($\mu$), the scalar expectation value ($S \mu$)
  and the QCD scale, while the second plot shows the same quantities
and the KK scale with the dependence on the radion factored out. The
vertical line shows the critical temperature $T_c \sim 700 \ \GeV$.
The dashed black line in the first plot shows the horizon location
$k^2 \rho_h$ as a function of temperature.}
\label{fig:MassScales}

\end{centering}	
\end{figure}

The minimum of the radion potential can be simply approximated in two
distinct regimes:
\begin{align}
  \mu(T)
  &=
  \left\{
  \begin{array}{ll}
    \mu_{\TeV} &  T < T_c \\
  \mu_{\TeV} \left( \frac{T}{cT_c} \right)^\frac{1}{1+\epsilon}   
  &  T \gg T_c
  \end{array}
  \right.
  \label{eq:radionvev} 
\end{align}
where the constant $c$ is given by:
\begin{align}
 	c^2 = \frac{4 v_\uv^2}{\epsilon^{3/2} v_\ir^2} \left(\frac{\mu_\TeV}{k}\right)^{2\epsilon}
\end{align}
The zero temperature value of the radion minimum $\mu_\TeV$ is well
approximated by equation~\ref{eq:RSmumin}, up to an 
$\mathcal{O} (\gamma v_s^3)$ correction to the $\mu^4$ co-efficient 
which is a result of the potential for $S$ not vanishing at the 
zero-temperature minimum.
The transition temperature $T_c$ is is the temperature at which the radion starts to move and is given by,
\begin{align}
  T_c^2
  &\simeq
  \frac{6}{N_s}
  \frac{m_\varphi^2}{m_s^2}
  \mu_\TeV^2
  =
  \frac{24\mu_\TeV^2}
  {N_s (\lambda_s v_s^2)}
  \epsilon^{3/2} v_\ir^2
  \label{eq:Tcestimate}
\end{align}
where $m_\varphi, m_s$ here are the zero-temperature masses for
$\varphi,s$. 
Since $\mu(T)/T$ is slowly growing, there is a maximum temperature
$T_{\rm max}$ beyond which the IR
brane would fall behind the horizon,
\begin{align}
  T_{\rm max}
  &\sim
  \mu(T_{\rm max})/\pi
  \Rightarrow
  T_{\rm max}
  =
 \frac{\mu_\TeV}{\pi} \left( \frac{\mu_\TeV}{\pi c T_c} \right)^{1/\epsilon
}
  \label{eq:Tmax}
\end{align}
This sets a (mild) bound on the reheat temperature of the universe.
This is at an exponentially high scale for 
$\delta \equiv \pi T_c/\mu_\TeV \ll 1/c$.
The AD transition temperature should also be higher
than the $Z_2$ symmetry restoration temperature, $T_c > T_s$.
These two requirements give us the following inequalities on our
parameter space,
\begin{align}
  \delta > \sqrt{\frac{6}{N_s}} \frac{m_\varphi}{m_s} > v_s
\end{align}
For illustration, we choose
the following benchmark  values\footnote{The value of $\epsilon$ was
chosen such that $\mu_\TeV \simeq 100\ \TeV$. A more generic choice
works equally well.}
\begin{align}
  \left\{
  k = 6 \times 10^{16}\GeV,
  \epsilon = 4.13\times 10^{-2},
  v_\uv = 10^{-3},
  v_\ir = 3 \times10^{-4},
\nonumber\right.\\\left.
  N_s = 100,
  \lambda_s = 1,
  v_s = 2 \times10^{-3},
  \gamma = -10^{-8}
\right\},
  \label{eq:num-params}
\end{align}
and find the following parameters
\begin{align}
  \mu_{\rm TeV } 
  &\simeq 1.8 \times 10^{-12} k \simeq 100 \ \TeV
  \\
  T_c
  &\simeq 
  700\ \GeV
  \,.
\end{align}
The maximum temperature $T_{\rm max}$ for this case is around $5\times 10^{11} \GeV$.
The temperature evolution of various scales in this benchmark is
illustrated in figure~\ref{fig:MassScales}.

Comparing the AD model to the usual RS model, the high temperature
behaviour is vastly different. In the RS model, as we move to high
temperature, thermal effects drive the IR brane towards the AdS boundary, 
where it eventually collapses to form an AdS-S
black hole~\cite{Hebecker2001}. The
model then remains in the black hole phase until tunnelling to the RS
phase through a first order phase transition which is highly
suppressed in calculable models, as was shown in
section~\ref{sec:Supercooled}. In contrast, the model
presented here describes a situation where the IR brane is stabilised
closer to the UV end of the warped direction at high temperatures,
then falls approximately linearly with temperature into the IR before
stabilising at a constant value deep in the IR as in the RS case. It
should be noted that at very high temperatures, the black hole phase
can still have a lower free energy than the AD phase, but the IR brane remains
meta-stabilised. The only way the system can transition to the
black hole phase is through a first order phase transition which is
exponentially suppressed.

The AD model also introduces a novel temperature dependence of the
mass scales on the IR brane, which scale linearly with the radion.
Assuming that the SM is confined to the IR brane, the left panel of
figure \ref{fig:MassScales} shows the dependence of the QCD scale with
temperature in the AD model. The other dimensional parameters of the
SM, such as the electroweak vacuum value, exhibit a similar scaling
above the critical temperature, $T_c$, of the theory. Above $T_c$ the
temperature only increases marginally relative to the other scales of
the theory. This leads to the KK modes being frozen out to arbitrarily
high temperature, as shown in the right panel of figure
\ref{fig:MassScales}. A similar behaviour would occur for any other
scales of the theory which are above $T_c$ -- they are frozen out to
much higher temperatures than in the usual RS model. In section
\ref{sec:Cosmology} we discuss this behaviour and its potential
implications for BSM phenomenology in more detail.

\section{Low Temperature Phenomenology}

\label{sec:ZeroTPheno}

In this section we show we study the constraints on the avoided
deconfinement model from collider results and ALP searches. For
simplicity we have assumed that
the SM is confined to the IR brane. More realistic models typically
have some or all SM fields propagating in the bulk, which can offer an
explanation of the hierarchical Yukawa couplings in the SM~\cite{Gherghetta2000,
Agashe2003, Casagrande2008} but also lead to new constraints from
flavour-violating processes~\cite{Agashe2005, Bauer2009}. The
constraints on the RS model have been well-studied. Here we will focus
on the new features that are required for the AD mechanism to work.

The qualitative features that the mechanism requires can be estimated
using the inequality in equation~\ref{eq:param-inequalities}. For
$\delta \ll 1$, we need, 
\begin{align}
  \frac{m_s}{\mu_\TeV} &\simeq v_s  < \delta 
  \nonumber
  \\
  \frac{m_\varphi}{\mu_\TeV} &<\delta^2 \sqrt{N_s}
  \,.
  \label{eq:param-estimate}
\end{align}
Experimental constraints on a very light radion will lead us to require
a large number of scalars, $N_s \gg 1$, with masses below the
confinement scale.

In
section~\ref{sec:effLagrangian} we write down the effective Lagrangian
-- the relevant degrees of freedom being the SM fields, the radion $
\varphi$, and the AD scalar(s) $s$. We ignore higher dimensional operators
which arise from integrating out KK modes and $1/N$ suppressed stringy
corrections, taking these to be negligible. In
section~\ref{sec:Experiment} we then describe the dominant
experimental constraints in different regions of parameter space.

\subsection{Effective Lagrangian at zero temperature}

\label{sec:effLagrangian}

The tree-level interactions of the radion $\varphi$ with the SM fields 
can be written compactly by the replacing the mass terms in
the SM by a $\varphi$ dependent mass,
\begin{align}
  \mathcal{L}^{(\rm tree)}[m_i] \to 
  \mathcal{L}^{(\rm tree)}_{\rm int} 
  \left[m_i\left(1 +\frac{\varphi}{F_\varphi} \right)\right]
\end{align}
where we the decay constant $F_\varphi$ was defined in
equation~\ref{eq:Fphi}.
This produces Yukawa-like interactions of $\varphi$ with the fermions,
as well as trilinear and quartic couplings with the Higgs and the
gauge bosons. This form of the radion potential is dictated by the AdS
isometries (and hence is valid in the limit of $\epsilon\ll 1$). The
self-interaction terms for the radion are generated by the GW mechanism,
\begin{align}
  \mathcal{L}_{\rm radion}
  &=
  \frac12 m_\varphi^2 \varphi^2 
  \left(
  1 
  + \frac53 \frac{\varphi}{F_\varphi} 
  + \frac{11}{12}\frac{\varphi^2}{F_\varphi^2}
  \right)
  \label{eq:rad-self}
\end{align}
where we have only kept terms to leading order in $\epsilon$. 
Finally, the scalars $s$ interact with the SM through the radion portal. 
\begin{align}
  \mathcal{L}^{(\rm tree)}_{\rm s}
  &=
  \lambda_s v_s^2 \mu_\TeV^2 s^2 \left(1+\frac{\varphi}{F_\varphi}\right)^2
  +\lambda_s v_s \mu_\TeV s^3 \left(1+\frac{\varphi}{F_\varphi}\right)
  +\frac14 \lambda_s s^4
  +\mathcal{O}(\gamma)
  \label{eq:Ls}
\end{align}
The terms suppressed by the explicit $Z_2$ violating
coupling $\gamma$ are assumed to be very small, and do not contribute
significantly to the zero-temperature phenomenology.
Notice that we do not generate terms of the form
of $\varphi$-s mixing, or $s \varphi^3$. The classical solution sets the
linear term in $s$ to zero, and the $\varphi$ field coupling as
$(1+\varphi)$ then does not have a linear coupling to $s$ around the
vacuum.
A small Higgs portal coupling of the form $\kappa s H^\dagger H$ can
be added in order for $s$ to be able to decay safely before BBN.

At loop level, there are also induced couplings between the EM and QCD
field strengths proportional to their $\beta$-functions:
\begin{align}
	\mathcal{L}^{(1 - \rm loop)}_{\rm int}  \supset
	 \frac{\alpha_{\rm EM} }{8\pi F_\varphi}
	b_{\rm EM} \varphi \,
	F_{\alpha \beta} F^{\alpha \beta}
	+ \frac{\alpha_{s} }{8\pi F_\varphi}
	b_{\rm G}  \varphi \,
	G^a_{\alpha \beta} G^{a \, \alpha \beta}  \, ,
	\label{eq:FSCouplings}
\end{align}
with the dominant contributions to these terms coming from quark and
$W$-boson loops. In the case where the SM is confined to the IR brane,
$b_{\rm EM} = 11/3$, $b_{\rm G} = -\frac{11}{3}N_c + 2n/3$, where $n$ is the
number of quarks lighter than the radion~\cite{Blum2014}.

\subsection{Experimental constraints}

\label{sec:Experiment}

The low energy phenomenology of the model is largely determined by the
physical masses of the radion and $s$ fields, as well as the KK
scale. These are related to the fundamental parameters of the model
by:
\begin{align}
    m_{\KK} &\simeq \frac{5 \pi}{4} \mu_\TeV	\, ,	\\
    m_{\varphi} &= 2\sqrt{2}v \epsilon^{3/4} \mu_\TeV	\, ,	\\
    m_{s} &= \sqrt{2\lambda_s} v_s \mu_\TeV
     \, .
\end{align}
Collider searches limit the KK scale in RS models to be above
$m_{\KK}\gtrsim 4.25$ TeV \cite{Sirunyan2018, Sirunyan2019}, requiring
the KK resonances to be out of the kinematic reach of current
colliders. Due to the approximate shift symmetry of the GW field
(broken only by the small parameter $\epsilon$), the radion is
parametrically lighter than the KK scale. The AD scalar masses are
similarly suppressed, with $m_{s}$ proportional to the
combination $(\lambda_s v_s^2)^{1/2}$, which is chosen to be small for
the $s$ phase transition to happen well before the deconfinement
transition.
Therefore, the radion and the AD scalars can be kinematically accessible at
colliders~\cite{Giudice2018}, however their couplings to the SM are
suppressed by $\mu_\TeV$. Collider constraints translate into a
bound $\mu_\TeV
\gtrsim 2 \, \TeV$~\cite{Blum2014}, which is weaker than direct
bounds on the KK scale.

If the radion mass is below the GeV scale,
bounds on the ${\varphi \gamma \gamma}$ coupling from supernova 
cooling\footnote{Whether supernova bounds on the radion coupling apply 
depends on the radion coupling to nucleons \cite{Abu-Ajamieh2017}. In the 
case where the SM quarks and gluons are on the IR brane, this coupling is 
too large for the radion to contribute significantly to supernova cooling.}, 
cosmology and beam dump experiments can give the strongest bounds on the 
model. These limits have been derived for axion-like
particles~\cite{Masso1995, Jaeckel2015, Dobrich2015}, which translate
into a bound
\begin{align}
  F_\varphi \gtrsim 4.25  \times 10^7 \, \TeV \, .
\end{align}
For a heavier radion $m_\varphi > 1 \GeV$, 
these constraints are no longer applicable. The radion mass will be
above a GeV for parameters,
\begin{align}
  v \epsilon^{3/4}& > 
  7 \times 10^{-5} \
  \left( \frac{\mu_{\TeV} }{5 \ \TeV} \right)^{-1}
  \, .
  \label{eq:RadionMassBound}
\end{align}
The AD scalars can couple to the SM fields through the radion portal.
For light AD scalars, the couplings to photons/gluons generated at
higher loop order might still provide significant constraints. If the
AD scalars are above the 1 GeV scale, these constraints are also
absent.

\section{Cosmology}

\label{sec:Cosmology}

The mechanism of avoided deconfinement has dramatic implications for
early universe cosmology. The main departure from a standard cosmology
is due to the scaling of the radion expectation value with
temperature.  This leads to the interesting consequence that while the
universe may be reheated to a very high temperature (exciting heavy
fields on the UV brane, for instance), from the IR brane
point-of-view, the cosmology resembles a low-reheat cosmology.  We aim
to highlight some of the applications of the AD model to cosmology,
but leave a detailed study of these implications for future work.

Figure \ref{fig:MassScales} shows the characteristic
dependence on temperature of dimensionful parameters on the IR brane.
 In particular,
$m_{KK} >T$ to arbitrarily high temperature, as required by the condition 
that the IR brane be stabilised outside the horizon at a given 
temperature. Therefore, KK modes play no role in 
early universe cosmology from the point of view of the IR brane. 
In the high temperature regime $T>T_c$, the radion expectation value scales with 
temperature as:
\begin{align}
	\mu(T) \propto T^{\frac{1}{1 + \epsilon}} \, .
\end{align}
This introduces a scaling of the other dimensionful quantities of the 
theory with $T$. The KK scale, the Higgs mass parameter and the QCD
scale and are all proportional to 
$\mu(T)$\footnote{More generally, $\Lambda_{\rm QCD}
\propto (\mu(T))^n$, where $n=1$ is true for the case where the SM is
confined to the IR 
brane, $n < 1$ if some of the SM quark fields are bulk fields.}, which means
that the ratio of the temperature to these mass scales (denoted
$\Lambda$) varies with $T$ as:
\begin{align}
	\frac{T}{\Lambda} \propto T^{\frac{\epsilon}{1+\epsilon}} \, .
\end{align}
The consequence of this scaling is that the ratio $T/\Lambda$ may reach unity 
at significantly higher temperatures than is the case in standard RS cosmology.
For example, as we show below, if the critical temperature $T_c$ is below the
electroweak symmetry breaking scale, then the electroweak symmetry
restoration phase transition
may occur at much higher temperatures than in the usual case, or never
occur at all.

\begin{figure}[tp]
  \begin{centering}
    \includegraphics[scale=0.25]{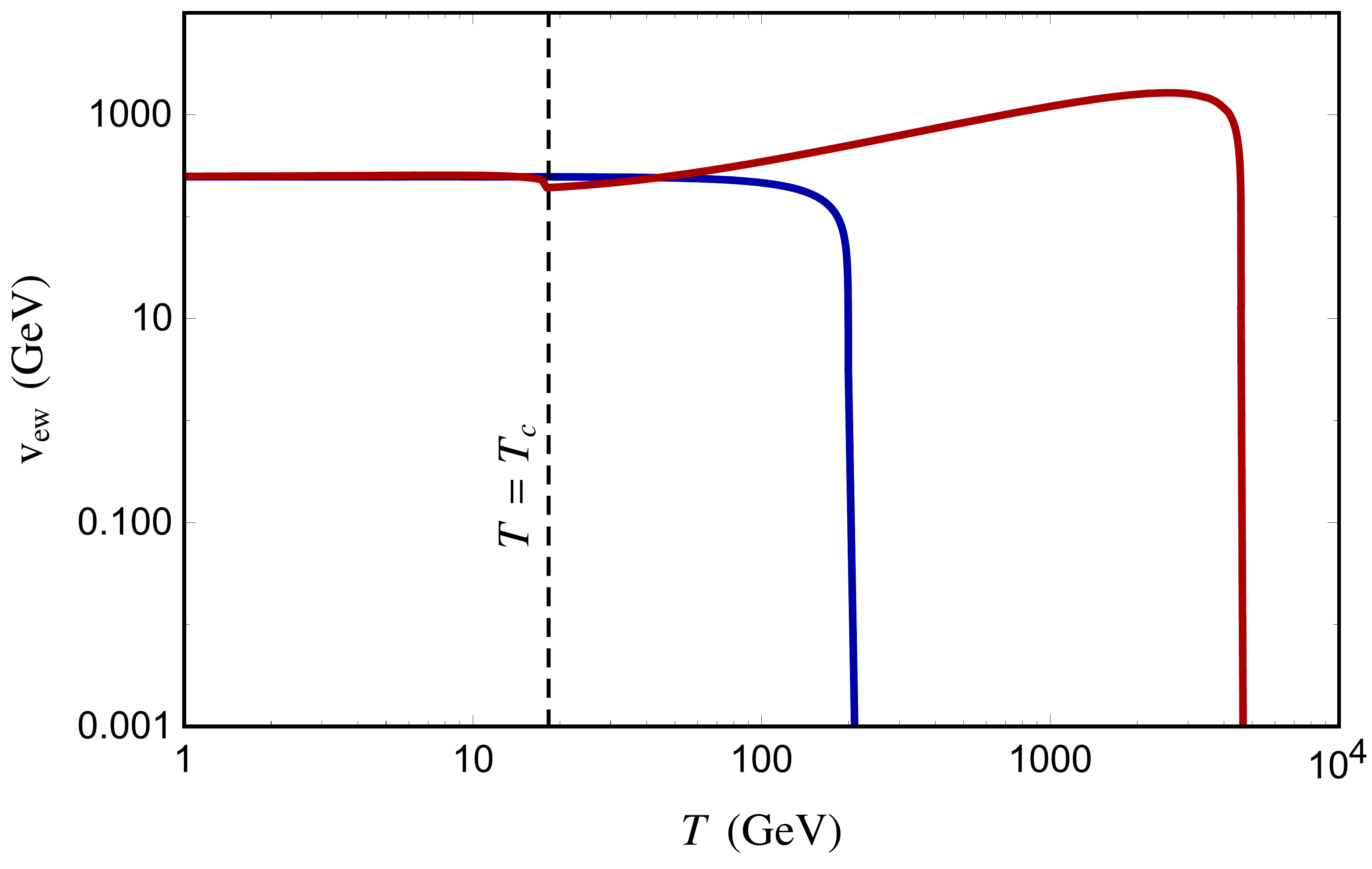}
    \caption{Higgs expectation value, $v_\ew$, as a function of 
    temperature for the parameters of equation~\eqref{eq:num-params} (blue) 
    and equation~\eqref{eq:num-params2} (red). The dashed line is the critical temperature, $T_c = 18 \GeV$ for the choice of parameters in equation~\eqref{eq:num-params2}.}
    \label{fig:Higgsmass}
  \end{centering}
\end{figure}

\subsection{Electroweak Phase Transition}
In this section we show that the electroweak phase
transition can occur in the avoided deconfinement phase at
temperatures much higher than the weak scale.
To illustrate some of these effects, we use a
new set of parameters, with:
\begin{align}
  v_\ir = 1.5 \times 10^{-4}, \,
  v_\uv = 7.5 \times 10^{-4}, \,
  \epsilon  =  0.05 \,
  \label{eq:num-params2}
\end{align}
and all other parameters as in equation~\eqref{eq:num-params},
which leads to a radion stabilised at $\mu_\TeV = 4.73$~TeV at zero 
temperature. We note that with these parameters, the model in its
simplest for does not
satisfy the bound on 
the radion mass~\eqref{eq:RadionMassBound}.
We expect that 
a more complete model with additional breaking of scaling invariance 
can lead to an unsuppressed radion mass and a less severe bound than 
equation~\eqref{eq:RadionMassBound}. This could happen, for example,
through additional terms in the GW action~\cite{Chacko:2014pqa}, 
allowing more fields to propagate in the bulk,
or by considering a more general geometry for the fifth 
dimension~\cite{Hassanain2007}. We will leave the detailed model
building for future work.

The Higgs potential at finite temperature gets thermal corrections
from the top Yukawa, gauge couplings and its quartic coupling. In
addition, the
Higgs mass parameter scales with $\mu(T)$.
The Higgs thermal mass in the low- and high-temperature limits is given
by:
\begin{align}
  \mu_h^2 (T>T_c) 
  &= 
  T^2  \left( -\lambda \frac{v_{\ew}^2}{c^2 T_c^2}
    \left(\frac{cT_c}{T}\right)^{2\epsilon}
    +
    \frac{\lambda_t^2}{4} + \frac{3 g^2}{16} + \frac{g'^2}{16} 
  + \frac{ \lambda}{2}  \right),
  \\
  \mu_h^2 (T<T_c) &= -\lambda v_{\ew}^2 + 
  T^2 \left( \frac{\lambda_t^2}{4} + \frac{3 g^2}{16} 
  + \frac{g'^2}{16} + \frac{\lambda}{2} \right).
\end{align}
where $v_{\rm ew} \simeq 246\GeV$. The electroweak phase transition
(EWPT)
happens at the temperature where the Higgs mass parameter become
positive. This happens for
$T<T_c$ if $T_c$ is above the
electroweak scale, in which case there is no modification to the phase
transition in comparison to the SM. However, if $T_c$ is below the
electroweak scale, the EWPT will occur at a temperature:
\begin{align}
	T_* =
	c T_c \left( \frac{T_{\ew}}{c T_c} \right)^{\frac{1}{\epsilon}}
\end{align}
where $T_{\ew}$ is the temperature of the EWPT in 
the SM. For small $\epsilon$, even a modest ratio $T_{\ew} / T_c$ can lead 
to the EWPT occurring at a temperature which is orders of 
magnitude above the scale predicted by the SM. Figure~\ref{fig:Higgsmass} 
shows $-\mu_h^2$ as a function of temperature the two sets of parameters 
defined in equations \eqref{eq:num-params} \& \eqref{eq:num-params2}.
For the second set of parameters the EWPT doesn't occur until 
the universe reaches a temperature of order $\sim 5 \times 10^3$ GeV.

A high temperature EWPT has been considered in refs~\cite{Meade2018,
Baldes:2018nel, Glioti:2018roy, Matsedonskyi2020} in the context of
electroweak baryogenesis. A primary motivation for these models is to
avoid the bounds which result from introducing new sources of CP
violation around the weak scale by having the EWPT occur at a
temperature $T \gg v_{\ew}$. This typically requires the introduction
of a large number of fields coupled to the Higgs sector in order to
significantly increase the temperature of the phase transition while
satisfying collider bounds. In contrast, the AD model provides a
mechanism in which the electroweak phase transition can occur at
arbitrarily high temperatures due solely to the Higgs-radion
interaction. However, the Higgs potential must still be modified to
make the EWPT first order and introduce new sources of CP violation
introduced in order to include a mechanism for electroweak
baryogenesis in the AD framework. Further, even though the
electroweak phase transition happens at much higher temperatures,
the scales governing local physics on the IR brane also scale
with $T$. Thus, if the CP violating operators are localized on the
IR brane, their effect at $T=T_*$ will be the same as that at
$T=T_c\lesssim v_{\rm ew}$.
Therefore, they will be subject to the very constraints the models
of~\cite{Meade2018, Baldes:2018nel, Glioti:2018roy, Matsedonskyi2020}
were constructed to avoid. On the other hand, if the CP violating
operators are on
the UV brane/bulk, their effect does become much more
important at higher temperatures. It will be interesting to construct
and study a high-temperature electroweak baryogenesis model using avoided
deconfinement in further detail.

\subsection{High Scale Baryogenesis}

As discussed in section~\ref{sec:Supercooled}, in the usual RS model
the high temperature phase is described by an AdS black hole, before a
phase transition to the RS phase at a temperature around the TeV scale
or below.  The period of supercooling accompanying the phase
transition significantly dilutes any pre-existing baryon
asymmetry~\cite{Baratella2019}.  This has motivated consideration of
baryogenesis mechanisms that combine the electroweak and RS phase
transitions~\cite{Bruggisser2018, Bruggisser2018a, Konstandin2011b}.
Baryogenesis mechanisms which operate at temperatures significantly
above the TeV scale are difficult to realise in the RS model. This is
not the case, however, for the AD model, as the universe is never in
the BH phase after inflation and does not undergo a period of
supercooling.

At high temperature the radion is stabilised closer to the UV brane. 
This means that fields localised toward the IR brane may have
significant overlap with UV- localised fields at early times, with the
UV and IR sectors then decoupling at low 
temperature. This allows for the possibility of having baryogenesis occur due to 
interactions between the IR and UV fields, which have $\mathcal{O}(1)$
couplings in the early universe but whose interactions are negligibly
small after the radion has settled to its 
zero-temperature expectation value.

\begin{figure}[tp]
  \begin{centering}
    \includegraphics[scale=0.8]{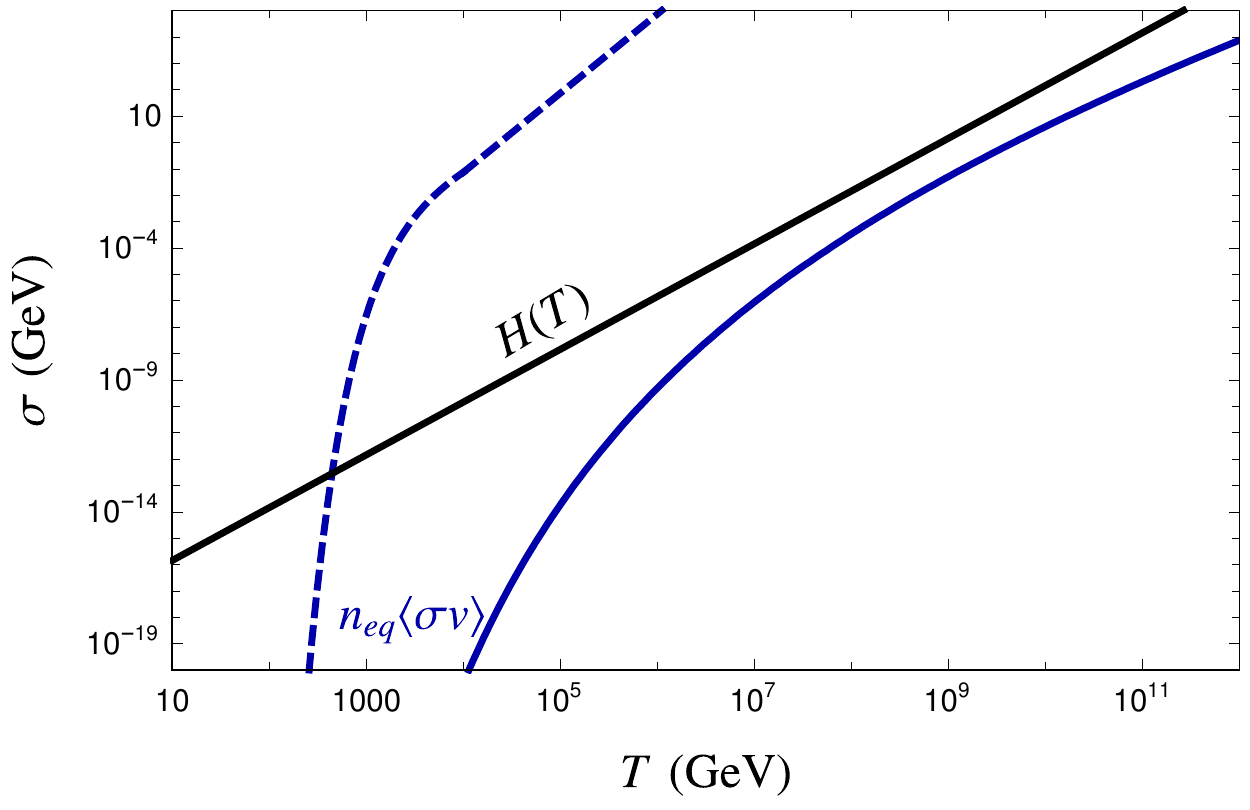}
    \caption{Dark matter annihilation rate for $m_\chi (T=0) = 10$ TeV, and $\alpha_{\rm{DM}} \sim 10^{-2}$ in the AD model (solid blue line) and without radion dependence (dashed blue line). The AD model parameters are as in equation~\eqref{eq:num-params2}. The Hubble rate assuming radiation domination and $g_* = 106.75$ is shown in black. }
    \label{fig:Freezeout}
  \end{centering}
\end{figure}

\subsection{WIMP Freeze-in}
The relic abundance of a particle with weak
scale mass and interactions in standard cosmology turns out to be
a good estimate for the observed dark matter abundance. This has led
to the WIMP paradigm, which is supported by the idea that new physics
at the weak scale is motivated by the electroweak hierarchy problem.
In this light, we expect a WIMP in the RS model to be associated with
a field localised to the IR brane. Thus, avoided deconfinement may
have significant
implications for the WIMP freeze-out in such cases -- in fact, we show
that particles with weak scale interactions can have a
freeze-in mechanism.

As noted above, the scales on the IR brane are proportional to $\mu
(T)$ and if the dark matter particle $\chi$ lives on the IR brane, the
quantity $m_\chi(T)/T$ changes by only an O(1) amount during the phase
of avoided deconfinement. In particular, the thermal abundance of $\chi$
can be Boltzmann suppressed for the entire cosmic history.

If the annihilation rate of dark matter is set by a weak coupling
$\alpha_{\rm{DM}}$, then the equilibrium annihilation rate in early cosmology
can be estimated as (assuming
$m_\chi (0) > T_c$),
\begin{align}
  n_{eq} \langle \sigma v \rangle
  &\sim
  \frac{\pi \alpha_{\rm{DM}}^2}{m_\chi(T)^2}
  (m_\chi(T) T)^{3/2} 
\exp \left[ -\frac{m_\chi(T)}{T} \right]
  \sim
  T
  \frac{\pi \alpha_{\rm{DM}}^2}{m_\chi(0)^2}
\exp \left[-\frac{m_\chi(0)}{cT_c} \right]
+O(\epsilon \log(T/T_c))
,
\end{align}
where we have ignored $\mathcal{O}$(1) factors to highlight the scaling behavior.
The form of the annihilation rate being nearly proportional to $T$ follows
from the approximate conformal symmetry. Figure \ref{fig:Freezeout} shows 
the different annihilation rates as a function of temperature in the AD model
and without radion dependence. 

The annihilation rate decreases slower than the Hubble rate, which
decreases as $T^2$ in radiation domination, as shown in 
figure~\ref{fig:Freezeout}. Even though the
zero-temperature annihilation rate is weak-scale, the high-temperature
annihilation rate can be out of equilibrium because it is Boltzmann
suppressed. After the IR brane is stabilized at $T \sim T_c$,
$m_\chi(T) \sim m_\chi(0)$, and the equilibrium annihilation rate
drops exponentially with temperature.  If the Hubble rate around $T =
T_c$ is larger than the equilibrium annihilation rate, this implies that:
\begin{align}
\frac{m_\chi(0)}{T_c}
\gtrsim
\log \frac{M_{\rm pl}}{T_c},
\end{align}
and the annihilation is never in thermal equilibrium, so the
abundance is set by freeze in. 
Note that the DM-SM coupling can be sizeable, so that we can
detect $\chi$ in direct and indirect detection experiments as a
WIMP. However the usual relic abundance calculation would not apply.
This can potentially open up a large parameter space for simple WIMPs
like the Wino-like electroweak triplet, or other heavy WIMP candidates
which would have a large abundance in the standard freeze out history.

\subsection{QCD Phase Transition and the QCD Axion}

The QCD phase transition in the AD model may also be modified from the
usual picture if $T_c$ is below the QCD scale. In order to achieve
this in our model while satisfying the bounds on the radion and $s$
masses requires a large number of AD scalars. However, as was the case
for the electroweak phase transition, even for $T_c$ slightly below
the QCD scale the QCD phase transition may occur at temperatures far
above the TeV scale. This may be able to reproduce some of the
non-standard QCD dynamics discussed in refs~\cite{Ipek:2018lhm,
Croon2020, Berger2020}.

A cosmology where QCD confinement occurs at high temperatures, when
$\Lambda_{\QCD} (T) \gg \Lambda_{\QCD}(0)$, can also have dramatic
consequences for the abundance of the QCD axion. The axion field can have various 5D 
origins; one simple possibility is
that it is the fifth component of $U(1)$ gauge field in 5D. Irrespective
of its origin, the large decay constant of the axion suggests that its
wavefunction is localised near the UV brane. Therefore, we can
safely assume that $f_a$ is largely temperature independent. 
In the confined phase of QCD, but with a temperature-dependent
confinement scale, the axion mass is given by,
\begin{align}
  m_a (T) = \frac{f_\pi (T) m_\pi(T)}{f_a}
  \label{eq:axion-mass}
\end{align}
The axion starts
oscillating around the epoch of QCD confinement
at a temperature $T_{\rm osc}$
which is defined by $m_a (T_{\rm osc}) \simeq H(T_{\rm osc})$.
The
axion abundance at the onset of oscillation is 
\begin{align}
  \rho_a (T_{\rm osc})
  &\sim
  m_a^2(T_{\rm osc}) f_a^2 \theta_i^2
\end{align}
where $\theta_i$ is the initial misalignment angle. 
The mass of the axion
continues to decrease due to the temperature dependence of
$\Lambda_{\QCD}(T)\sim \mu(T)$. In the adiabatic approximation
$\dot{m}\ll m^2$,
the number density of the axion scales as $a^{-3}$, and the mass
redshifts as $\sim a^{-2}$, so the axion
energy density redshifts approximately as $a^{-5}$ in this epoch,
whereas the background energy density is redshifting as $a^{-4}$. 
This can reduce the axion abundance dramatically.

\subsection{Gravitational Waves}

The absence of a first order confinement phase transition is a
necessary feature of this mechanism that distinguishes it from the
standard RS model. The RS phase transition results in a gravitational
wave signal which will be absent in the avoided deconfinement
model~\cite{Megias2018a, Randall2006}. The RS phase transition also
leads to a drop in $g_*$ of order $N^2$ as a result of degrees of
freedom confining and freezing out. If there is an observable
background of gravitational waves, such as from a cosmic string
network~\cite{Cui2017, Cui2018} or inflation~\cite{Watanabe2006,
Jinno2012, Saikawa2018},  this change in $g_*$ is observable as a
relative decrease in the power in modes which were below the horizon
scale prior to the phase transition. The absence of these
gravitational wave signals could be used to distinguish the AD model
from RS models which do undergo a phase transition. Furthermore, the
addition of the AD scalars, which are necessarily light degrees of
freedom due to the bound \eqref{eq:param-estimate}, also leads to a
potentially observable change in $g_*$ in the early universe for $N_s$
as low as $N_s \sim 10$ and masses around the GeV scale.

In addition to modifying the RS phase transition, in our set up there
are additional phase transitions associated with the $s$ fields, which
can be first order and can each happen at slightly different temperatures.
This can give us interesting forest of GW signals with a spectrum that is
different from the one expected from a single phase transition. As
noted above, the electroweak and/or the QCD phase transition may also
be made first order and can happen at very high temperatures,
predicting a gravitational wave signature from these phase transitions
as well.

\section{Discussion}

\label{sec:Discussion}

In this work we have described a mechanism which addresses the
cosmological problem of eternal inflation due to suppressed
 confinement transitions in the RS model. 
The standard RS model is described at high
temperature by an AdS black hole, with a transition to the RS phase
proceeding via a first order phase transition which is exponentially
suppressed by the large number $N^2$. We have shown that this
situation can be avoided by introducing new scalars localised in the
IR which generate a potential that stabilises the IR brane at high
temperatures. Provided the universe exits inflation in the RS phase,
it remains there, never entering the BH phase.

There are a number of issues that would be worth exploring
further.
It would be interesting to understand this phenomenon in a 4D field
theory example. The additional scalars that we have introduced on the
IR brane are expected to be emergent degrees of freedom in the 4D
theory that appear after confinement. In such an example the thermal
effect of these scalars will be to drive the confinement scale itself
to higher values. This may provide a new insights on the problem of
confinement.

There are also various phenomenological applications of this
mechanism which we have merely touched upon in this paper. Avoided
Deconfinement changes the cosmological history in a unique way, where
from the IR brane point of view it is a model with effectively a low reheat
temperature, but from the UV brane point of view the temperature can
get arbitrarily high. This allows us to build realizations where the
electroweak and/or QCD phase transitions happen at very
high temperatures. Since the cosmology at high temperatures is
modified, we have shown that the mechanism for generating the
abundance of various species such as WIMP dark matter, axion dark
matter or baryogenesis can be significantly modified. An
interesting future direction would be to build explicit 
models which realize these mechanisms, and study their
phenomenological signatures. 

Gravitational waves are a powerful experimental tool for studying very early
universe cosmology, both in terms of new sources of the waves as well
as modifications of propagation of gravitational waves in the early
universe plasma.  Modification of the Randall-Sundrum phase
transition, or the electroweak/QCD phase transition can change the
expectations of GW signals from these
phase transitions; these phase transitions are also associated with a
change of number of degrees of freedom in the plasma, which may also
be detectable in the GW spectrum. Even if a large-$N$ confining gauge group is
part of a dark sector decoupled from the standard model, these
gravitational wave signatures can provide important information about
these sectors.  Thus, the detailed phenomenological predictions of avoided
deconfinement would be important to study further even in this more
general situation.

\begin{acknowledgments}
  We would like to thank Raman Sundrum and Soubhik Kumar for useful
  discussions and comments on the manuscript. We are grateful to Anson
  Hook, Lisa Randall, Matt Reece and John March-Russell for
  useful conversations. 
  PA is supported by the STFC under Grant No. ST/T000864/1.
  MN is funded by a joint Clarendon and Sloane-Robinson scholarship
  from Oxford University and Keble college.
\end{acknowledgments}

\appendix

\section{Bounce action for deconfining phase transition}

\label{app:tunneling}

In this appendix we estimate the bounce action, $B$, which determines the transition rate from the AD phase in the high-temperature regime to the AdS-S or deconfined phase. At high temperatures the phase transition from the AD phase to the black hole phase proceeds at a rate
\begin{equation}
	\Gamma \simeq T^4 e^{-B} .
\end{equation}
If this is larger than $H^4$, where $H$ is the hubble rate, then this indicates that the AD phase is unstable. This defines a maximum temperature $T_{\rm max}$ above which the AD mechanism no longer works, but we will find that tunnelling only becomes significant at temperatures equal to the temperature which defines the classical instability of the model (defined in equation~\eqref{eq:Tmax}), up to $\mathcal{O}(1/N)$ corrections.

In order to determine $B$ we make the approximation that the action is dominated by the dynamics of the radion and neglect the contribution from the gravitational portion of the action. The justification for this is that the gravitational action scales as $N^2$ with no further enhancement from small or large parameters, while the contribution to the bounce from the radion, as we show below, scales like $N^2 \lambda^{-1/2}$ for a weak coupling $\lambda$. In this approximation the relevant Euclidean action is
\begin{equation}
\begin{aligned}
	S_{E} &= \frac{N^2}{4 \pi} \int_0^{T^{-1}} d t_E \int r^2 dr 
	\left[ (\partial  \mu)^2 
	-\lambda_1  T^2 \mu^2 
	+ \lambda_2  \mu^4\right],	\\	
	\lambda_1 &= \frac{ 2 \pi^2 N_s \lambda_s v_s^2}{3 N^2}	,	\\
	\lambda_2 &= \frac{64\pi^2 (v_\ir - v_\uv)^2}{N^2},
\end{aligned}	
\end{equation}
where we have explicitly scaled out the factor of $N^2$. At high temperature the minimum of the radion potential is well-approximated by equation \eqref{eq:radionvev} 
\begin{equation}
	\mu \simeq \alpha (T) T,  \qquad
	\alpha (T) = 
	\frac{\mu_\TeV }{(c T_c T^\epsilon)^{\frac{1}{1+\epsilon}}}.
\end{equation}
After rescaling the co-ordinates and radion field as $\mu = \alpha T \tilde \mu$, $x_E = T^{-1} \tilde x_E$ the action can be written as:
\begin{align}
	S_{E} &= \frac{\alpha^2 N^2}{4 \pi} \int_0^1 d \tilde t_E 
	\int \tilde  r^2 d \tilde r 
	\left[ (\tilde  \partial \tilde  \mu)^2 
	-\lambda_1 \tilde  \mu^2 
	+ \lambda_2 \alpha^2 \tilde \mu^4\right]	.
\end{align}

\begin{figure}[tp]
\begin{centering}

  \includegraphics[width=0.45\textwidth]{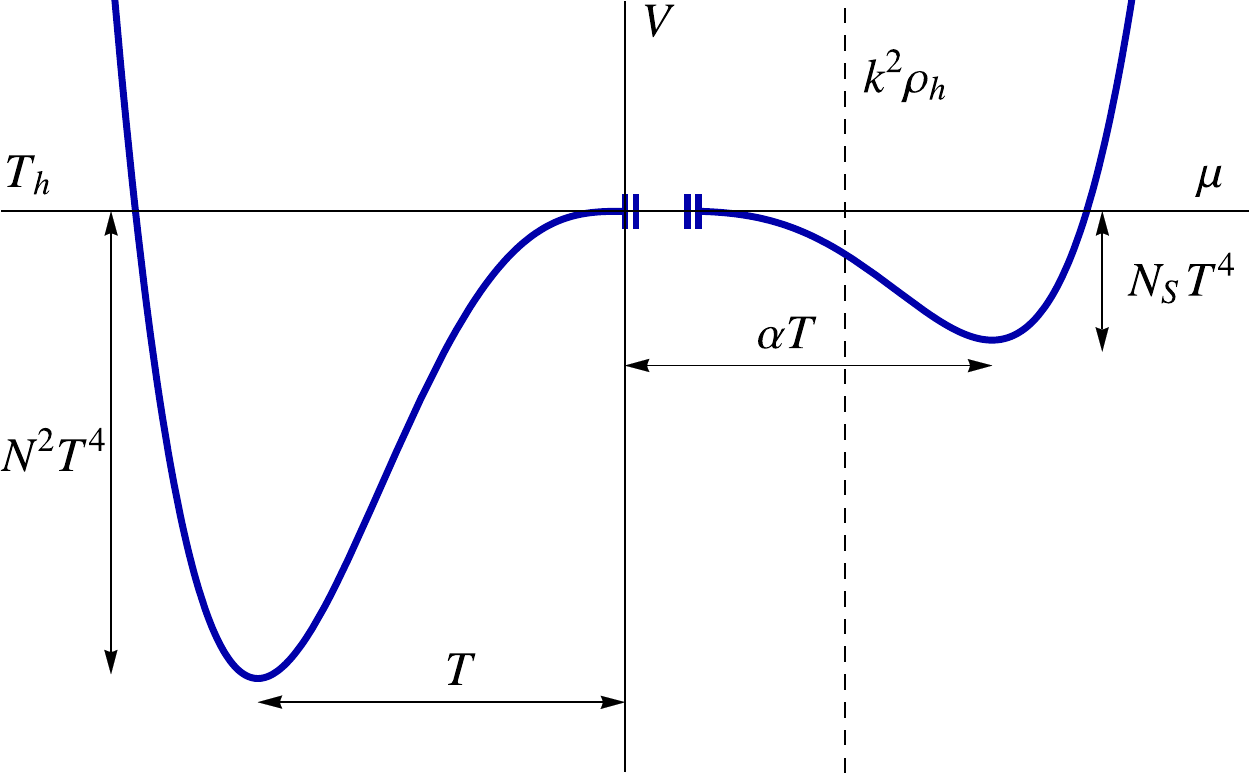}

\end{centering}

\caption{ Plot showing the parametric dependence of the potential in the AD model. The right-hand side shows the radion potential at high temperature, with depth set by $N_s T^4$ and width set by $\langle \mu \rangle =  \alpha(T) T$ and the left-hand side is the potential for the black hole hawking temperature $T_h$, with width set by $T$ and depth of order $N^2 T^4$. The dashed line shows the would-be horizon position on the radion side of the potential and the break in the curve indicates the region where EFT control is lost.}

\label{fig:tunnellingpotential}
\end{figure}

The equation of motion for $\tilde \mu$ then implies that $\tilde \mu$ varies by an $\mathcal{O}(1)$ amount over a distance of order $\Delta \tilde r \sim \lambda_1^{-1/2} \gg 1$, i.e.
\begin{align}
	\left | \frac{\partial \tilde \mu}{\partial \tilde r}  
	\right | \sim  \lambda_1^{1/2} .
\end{align}
We then make the conservative estimate that the bounce solution only requires the radion to vary by an amount given by
\begin{align}
	\Delta \mu = \alpha (T) T - k^2 \rho_h ,
\end{align}
which amounts to moving the IR brane from its stabilised location to the position of the would-be horizon. In figure~\ref{fig:tunnellingpotential} this corresponds to the radion varying from its value at the minimum of the potential to the dashed line, as opposed to a bounce analogous to the one proposed in~\cite{Creminelli2002} which involves the radion varying to $\mu =0$ over the bounce trajectory. In terms of $\tilde \mu$ this is
\begin{align}
	\delta \tilde \mu = 1 - \frac{\pi}{\alpha},
\end{align}
which approaches $0$ logarithmically (indicating that the IR brane is becoming classically unstable) as $T$ approaches $T_{\rm max}$.

With this estimate the characteristic radius of the tunnelling configuration will be $\tilde R_b \sim (\delta \tilde\mu) \lambda_1^{-1/2}$. For $\tilde R_b \gg 1$ we expect the bounce solution to be the $O(3)$ symmetric ($\tilde t_E$-independent) configuration, while for $\tilde R_b \ll 1$ the solution will obey an $O(4)$ symmetry and depend on the variable $ \tilde \rho = \sqrt{\tilde t_E^2 + \tilde r^2}$. The $O(4)$ bounce solution therefore only becomes dominant for $\delta\tilde\mu \ll \lambda^{1/2} \sim 1/N$ at which point the IR brane is close to becoming classically unstable anyway, so only the $O(3)$ symmetric bounce is relevant for computing the lifetime of the AD phase. The integrals determining the bounce action scale as
\begin{equation}
\begin{aligned}
	\int \tilde  r^2 d \tilde r 
	\left( (\tilde  \partial \tilde  \mu)^2 
	-\lambda_1 \tilde  \mu^2 \right)	 &= \lambda_1^{-1/2}  c_1  				\\
	  \int \tilde  r^2 d \tilde r \left( \lambda_2 \alpha^2 \tilde \mu^4\right)	&= \lambda_1^{-3/2} \lambda_2 c_2	.
\end{aligned}		  
\end{equation}
 where $c_1,c_2$ are $\mathcal{O}(1)$ coefficients. We can then estimate the bounce to be:
\begin{equation}
\begin{aligned}
	B &\simeq \frac{\alpha^2 N^2 (\delta 
	\tilde\mu)^3 \lambda_1^{-3/2} }{4 \pi} 
	\left[ c_1 \lambda_1 
	+ c_2 \lambda_2 \alpha^2 \right]	,			\\
	&\simeq \frac{ \sqrt{3} \alpha^2  N^3 ( \delta \tilde \mu)^3}
	{4 \sqrt{2} \pi^2 \left( N_s \lambda_s v_s^2\right)^{1/2}}
	\left[  c_1
	+ c_2  \frac{ 96 \alpha^2 (v_\ir - v_\uv)^2}{ N_s \lambda_s v_s^2}    \right].
\end{aligned}	
\end{equation}
The bounce action is therefore suppressed by the factor $N^3$ and additionally by inverse powers of $v_s$. The lifetime of the metastable AD vacuum is much larger than Hubble for temperatures up until $\delta \tilde \mu \lesssim \mathcal{O} (1/N)$, meaning we can safely ignore the tunneling rate from the AD to deconfined phase and consider only the classical instability of the model.

\bibliographystyle{JHEP}
\bibliography{ref.bib}

\end{document}